\documentclass[11pt]{article}

\usepackage{geometry}
\usepackage{theorem}
\usepackage{graphicx}
\usepackage{amsmath}
\usepackage{float}
\usepackage{amssymb}

\date{}
\newtheorem{theorem}{Theorem}
\newtheorem{definition}{Definition}
\newtheorem{lemma}{Lemma}

\newtheorem{specification}{Specification}
\newtheorem{proposition}{Proposition}

\newenvironment{proof}{{\noindent\bf Proof. } }{{\hfill $\Box$}}
\newenvironment{sketchproof}{{\noindent\bf Sketch of proof. } }{{\hfill $\Box$}}

\floatstyle{ruled}
\newfloat{algorithm}{tbp}{loa}
\floatname{algorithm}{Algorithm}

\begin{document}


\newcommand{\AN}{$\mathcal{SSMFP}_{1}$~}
\newcommand{\AD}{$\mathcal{SSMFP}_{2}$~}


\title{Two snap-stabilizing point-to-point communication\\ protocols in message-switched networks}

\author{
Alain Cournier\thanks{MIS Laboratory, Universit\'e de Picardie Jules Verne, alain.cournier@u-picardie.fr}
\and 
Swan Dubois\thanks{LIP6 - UMR 7606 Universit\'e Pierre et Marie Curie - Paris 6 \& INRIA Rocquencourt, swan.dubois@lip6.fr}
\and
Vincent Villain\thanks{MIS Laboratory, Universit\'e de Picardie Jules Verne, vincent.villain@u-picardie.fr}
}

\maketitle


\begin{abstract}
A \emph{snap-stabilizing} protocol, starting from any configuration, always behaves according to its specification. In this paper, we present a snap-stabilizing protocol to solve the message forwarding problem in a message-switched network. In this problem, we must manage resources of the system to deliver messages to any processor of the network. In this purpose, we use information given by a routing algorithm. By the context of stabilization (in particular, the system starts in an arbitrary configuration), this information can be corrupted. So, the existence of a snap-stabilizing protocol for the message forwarding problem implies that we can ask the system to begin forwarding messages even if routing information are initially corrupted.

In this paper, we propose two snap-stabilizing algorithms (in the \emph{state model}) for the following specification of the problem:
\begin{itemize}
	\item Any message can be generated in a finite time.
	\item Any emitted message is delivered to its destination once and only once in a finite time.
\end{itemize}
This implies that our protocol can deliver any emitted message regardless of the state of routing tables in the initial configuration.

These two algorithms are based on the previous work of \cite{MS78}. Each algorithm needs a particular method to be transform into a snap-stabilizing one but both of them do not introduce a significant overcost in memory or in time with respect to algorithms of \cite{MS78}.
\end{abstract}

\newpage

\section{Introduction}

The quality of a distributed system depends on its \emph{fault-tolerance}. Many fault-tolerant schemes have been proposed. For instance, \emph{self-stabilization} (\cite{D74}) allows to design a system tolerating arbitrary transient faults. A self-stabilizing system, regardless of the initial state of the system, is guaranteed to converge into the intended behavior in a finite time. An other paradigm called \emph{snap-stabilization} has been introduced in \cite{BDPV07,BDPV99}. A snap-stabilizing protocol guarantees that, starting from any configuration, it always behaves according to its specification. In other words, a snap-stabilizing protocol is a self-stabilizing protocol which stabilizes in 0 time unit.

In a distributed system, it is commonly assumed that each processor can exchange messages only with its \emph{neighbors} (\emph{i.e.} processors with which it shares a communication link) but processors may need to exchange messages with \emph{any} processor of the network. To perform this goal, processors have to solve two problems: the determination of the path which messages have to follow in the network to reach their destinations (it is the \emph{routing} problem) and the management of network resources in order to forward messages (it is the \emph{message forwarding} problem).

These two problems received a great attention in literature. The routing problem is studied for example in \cite{BLT93,CM82,FG97,FG98,G00,LT87,LT94,MS79,T77,T80b} and self-stabilizing approach can be found (directly or not) in \cite{HC92,KK05,D97,JT03}. The forwarding problem has also been well studied, see \cite{D96,MS78,SJ95,T80,TS81,TU81} for example. As far we know, the message forwarding problem was never directly studied with a snap-stabilizing approach (note that the protocol proposed by \cite{JT03} can be used to perform a self-stabilizing forwarding protocol for dynamic networks since it is guaranteed that routing tables remain loop-free even if topological changes are allowed).

Informally, a message forwarding protocol allows any processor of the network to send messages to any destination of the network knowing that a routing algorithm computes the path that messages have to follow to reach their destinations. Problems come of the following fact: messages traveling through a \emph{message-switched network} (\cite{T01}) must be stored in each processor of their path before being forwarded to the next processor on this path. This temporary storage of messages is performed with reserved memory spaces called buffers. Obviously, each processor of the network reserves only a \emph{finite} number of buffers for the message forwarding. So, it is a problem of bounded resources management which exposes the network to deadlocks and livelocks if no control is performed. 

In this paper, we focus on message forwarding protocols which deal the problem with a snap-stabilizing approach. The goal is to allow the system to forward messages (without looses) regardless of the state of the routing tables. Obviously, we need that theses routing tables repair themselves in a finite time. So, we assume the existence of a self-stabilizing protocol to compute routing tables (see \cite{HC92,KK05,D97}).

In the following, a \emph{valid} message is a message which has been generated by a processor. As a consequence, an \emph{invalid} message is a message which is present in the initial configuration. We can now specify the problem. We propose a specification of the problem where message duplications (\emph{i.e.} the same message reaches its destination many time while it has been generated only once) are forbidden:

\begin{specification}[$\mathcal{SP}$]\label{specif:spe2}
Specification of message forwarding problem forbidding duplication.
\begin{itemize}
	\item Any message can be generated in a finite time.
	\item Any valid message is deliver to its destination once and only once in a finite time.
\end{itemize}
\end{specification}

In this paper, we investigate the possibility to transform two known message forwarding protocols (\cite{MS78}) into snap-stabilizing ones. We use a different scheme for both of them but we prove that these two schemes do not significantly modify time and space complexities of these protocols. Consequently, the main contribution of this paper is to show that it is possible to provide stronger safety properties without significant overcost.

The sequel of this paper is organized as follows: we present first our model (section \ref{sec:Model}). We quickly survey the seminal work of \cite{MS78} in section \ref{sec:survey}. Then we give, prove, and analyze our two solutions (sections \ref{sec:protocolN} and \ref{sec:protocolD}). Finally, we conclude by some remarks and open problems (section \ref{sec:Conclusion}).

\section{Model and definitions}\label{sec:Model}

We consider a network as an undirected connected graph $G=(V,E)$ where $V$ is a set of processors and $E$ is the set of bidirectional asynchronous communication links. In the network, a communication link $(p,q)$ exists if and only if $p$ and $q$ are \emph{neighbors}. Every processor $p$ can distinguish all its links. To simplify the presentation, we refer to a link $(p,q)$ of a processor $p$ by the label $q$. We assume that the labels of $p$ are stored in the set $N_{p}$. 

We also use the following notations: respectively, $n$ is the number of processors, $\Delta$ the maximal degree, and $D$ the diameter of the network. If $p$ and $q$ are two processors of the network, we denote by $dist(p,q)$ the length of the shortest path between $p$ and $q$ (\emph{i.e.} the \emph{distance} between $p$ and $q$). In the following, we assume that the network is \emph{identified}, \emph{i.e.} each processor have an identity which is unique on the network. Moreover, we assume that all processors know the set $I$ of all identities of the network.

\subsection{State model}\label{sub:State-model}

We consider the classical \emph{local shared memory model} of computation (see \cite{T01}) in which communications between neighbors are modeled by direct reading of variables instead of exchange of messages. 

In this model, the program of every processor consists in a set of \emph{shared variables} (henceforth, referred to as \emph{variables}) and a finite set of \emph{actions}. A processor can write to its own variables only, and read its own variables and those of its neighbors. Each action is constituted as follows: $< label >::< guard >\longrightarrow< statement >$. The \emph{label} is a name to refer to the rule in the discussion. The \emph{guard} of an action in the program of $p$ is a Boolean expression involving variables of $p$ and its neighbors. The \emph{statement} of an action of $p$ updates one or more variables of $p$. An action can be executed only if its guard is satisfied.

The \emph{state} of a processor is defined by the value of its variables. The state of a system is the product of the states of all processors. We refer to the state of a processor and the system as a (local) \emph{state} and (global) \emph{configuration}, respectively. We note $\mathcal{C}$ the set of all configurations of the system. 

Let $\gamma\in\mathcal{C}$ and $A$ an action of $p$ ($p\in V$). $A$ is \emph{enabled} for $p$ in $\gamma$ if and only if the guard of $A$ is satisfied by $p$ in $\gamma$. Processor $p$ is \emph{enabled} in $\gamma$ if and only if at least one action is enabled at $p$ in $\gamma$. Let a distributed protocol $\mathcal{P}$ be a collection of actions denoted by $\rightarrow$, on $\mathcal{C}$. An \emph{execution} of a protocol $\mathcal{P}$ is a maximal sequence of configurations $\Gamma=\gamma_{0}\gamma_{1}...\gamma_{i}\gamma_{i+1}...$ such that, $\forall i\geq0,\gamma_{i}\rightarrow\gamma_{i+1}$ (called a \emph{step}) if $\gamma_{i+1}$ exists, else $\gamma_{i}$ is a terminal configuration. \emph{Maximality} means that the sequence is either finite (and no action of $\mathcal{P}$ is enabled in the terminal configuration) or infinite. All executions considered here are assumed to be maximal. $\mathcal{E}$ is the set of all executions of $\mathcal{P}$.

As we already said, each execution is decomposed into steps. Each atomic step is composed of three sequential phases: (i) every processor evaluates its guards, (ii) a \emph{daemon} chooses some enabled processors, (iii) each chosen processor executes one of its enabled actions. When the three phases are done, the next step begins. A daemon can be defined in terms of \emph{fairness} and \emph{distribution}. There exists several kinds of fairness assumption. Here, we present the \emph{strong fairness}, \emph{weak fairness}, and \emph{unfairness} assumptions. Under a \emph{strongly fair} daemon, every processor that is enabled infinitely often is chosen by the daemon infinitely often to execute an action. When a daemon is \emph{weakly fair}, every continuously enabled processor is eventually chosen by the daemon. Finally, the \emph{unfair} daemon is the weakest scheduling assumption: it can forever prevent a processor to execute an action except if it is the only enabled processor. Concerning the distribution, we assume that the daemon is \emph{distributed} meaning that, at each step, if one or several processors are enabled, then the daemon chooses at least one of these processors to execute an action. 

We consider that any processor $p$ is \emph{neutralized} in the step $\gamma_{i}\rightarrow\gamma_{i+1}$ if $p$ was enabled in $\gamma_{i}$ and not enabled in $\gamma_{i+1}$, but did not execute any action in $\gamma_{i}\rightarrow\gamma_{i+1}$. To compute the time complexity, we use the definition of \emph{round} (introduced in \cite{DIM97} and modified by \cite{BDPV07}). This definition captures the execution rate of the slowest processor in any execution. The first round of $\Gamma\in\mathcal{E}$, noted $\Gamma'$, is the minimal prefix of $\Gamma$ containing the execution of one action or the neutralization of every enabled processor from the initial configuration. Let $\Gamma''$ be the suffix of $\Gamma$ such that $\Gamma=\Gamma'\Gamma''$. The second round of $\Gamma$ is the first round of $\Gamma''$, and so on.

\subsection{Message-switched networks}\label{sub:Message-switched-network}

Today, most of computer networks use a variant of the \emph{message-switching} method (also called \emph{store-and-forward} method). It is why we have choose to work with this switching model. In this section, we are going to present this method (see \cite{T01} for a detailed presentation).

Each processor has $\mathcal{B}$ buffers for temporarily storing messages. The model assumes that each buffer can store a whole message and that each message needs only one buffer to be stored. The switching method is modeled by four types of moves:

\begin{enumerate}
\item \textbf{Generation}: when a processor sends a new message, it ``creates'' a new message in one of its empty buffers. We assume that the network may allow this move as soon as at least one buffer of the processor is empty. 
\item \textbf{Forwarding}: a message $m$ is forwarded (copied) from a processor $p$ to an empty buffer in the next processor $q$ on its route (determined by the routing algorithm). We assume that the network may allow this move as soon as at least one buffer buffer of the processor is empty. 
\item \textbf{Consumption}: A message $m$ occupying a buffer in its destination is and delivered to this processor. We assume that the network may always allow this move. 
\item \textbf{Erasing}: a message $m$ is erased from a buffer. We assume that the network may allow this move as soon as the message is forwarded at least one time or delivered to its destination.
\end{enumerate}

\subsection{Stabilization}\label{sub:Stabilization}

In this section, we give formal definitions of self- and snap-stabilization using notations introduced in \ref{sub:State-model}.

\begin{definition} [Self-Stabilization \cite{D74}] \label{def:self}
Let $\mathcal{T}$ be a task, and $\mathcal{\mathcal{S}_{T}}$ a specification of $\mathcal{T}$. A protocol $\mathcal{P}$ is self-stabilizing for $\mathcal{S_{T}}$ if and only if $\forall\Gamma\in\mathcal{E}$, there exists a finite prefix $\Gamma'=(\gamma_{0},\gamma_{1},...,\gamma_{l})$ of $\Gamma$ such that any executions starting from $\gamma_{l}$ satisfies $\mathcal{S_{T}}$.
\end{definition}

\begin{definition} [Snap-Stabilization \cite{BDPV99,BDPV07}] \label{def:snap}
Let $\mathcal{T}$ be a task, and $\mathcal{\mathcal{S}_{T}}$ a specification of $\mathcal{T}$. A protocol $\mathcal{P}$ is snap-stabilizing for $\mathcal{S_{T}}$ if and only if $\forall\Gamma\in\mathcal{E}$, $\Gamma$ satisfies $\mathcal{S_{T}}$.
\end{definition}

This definition has the two following consequences. We can see that a snap-stabilizing protocol for $\mathcal{S_{T}}$ is a self-stabilizing protocol for $\mathcal{S_{T}}$ with a stabilization time of 0 time unit. A common method used to prove that a protocol is snap-stabilizing is to distinguish an action as a ``starting action'' (\emph{i.e.} an action which initiates a computation) and to prove the following property for every execution of the protocol: if a processor requests it, the computation is initiated by a starting action in a finite time and every computation initiated by a starting action satisfies the specification of the task. We use these two remarks to prove snap-stabilization of our protocol in the following of this paper.

\section{Fault-free protocols}\label{sec:survey}

In this section, we survey the seminal work of \cite{MS78}\footnote{The reader is referred to \cite{T01} to find a much detailed description of this work.}. Remind that this work assume that routing tables are correct in the initial configuration.  To simplify the presentation, we assume that the routing algorithm induces only minimal paths in number of edges.

We have seen in section \ref{sub:Message-switched-network} that, by default, the network always allows message moves between buffers. But, if we do no control on these moves, the network can reach unacceptable situations such as \emph{deadlocks}, \emph{livelocks} or \emph{message losses}. If such situations appear, specifications of message forwarding are not respected.

In order to avoid deadlocks, we must define an algorithm which permits or forbids various moves in the network (functions of the current occupation of buffers). A such algorithm is a \emph{controller}. If a controller $\mathcal{C}$ ensure the following property: in any execution, $\mathcal{C}$ prevents the network to reach a deadlock, then $\mathcal{C}$ is a \emph{deadlock-free} controller.

Livelocks can be avoided by fairness assumptions on the controller for the generation and the forwarding of messages. Message losses are avoided by the using of identifier on messages. For example, one can use the concatenation of the identity of the source and a two-value flag in order to distinguish two consecutive identical messages generated by the same processor for a Destination $d$ (since all messages follow the same path).

Then, a deadlock-free controller which prevents also livelocks and message losses satisfies the specification of the message forwarding problem.

In the case where routing table are initially correct, \cite{MS78} introduced a generic method to design deadlock-free controllers. It consists to restrict moves of messages along edges of an oriented graph $BG$ (called \emph{buffer graph}) defined on the network buffers. Then, it is easy to see that cycles on $BG$ can lead to deadlocks. So, authors show that, if $BG$ is acyclic, they can define a deadlock-free controller on this buffer graph. In the sequel of this section, we present the two buffer graph which we use in our snap-stabilizing protocols.

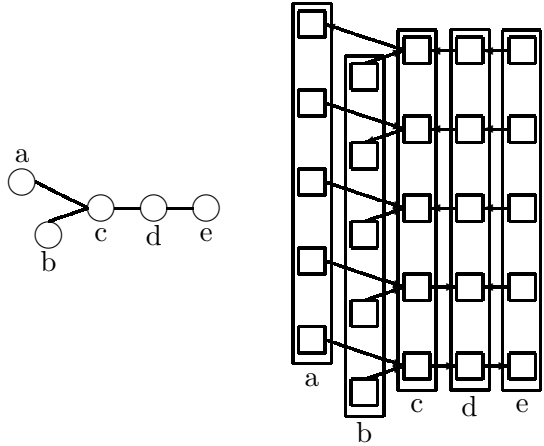
\begin{figure}
\noindent \begin{centering}
\ifx\JPicScale\undefined\def\JPicScale{0.35}\fi
\unitlength \JPicScale mm
\begin{picture}(216.97,165)(0,0)
\linethickness{0.3mm}
\put(20,97.12){\circle{10}}

\linethickness{0.3mm}
\put(50,87.12){\circle{10}}

\linethickness{0.3mm}
\put(30,77.12){\circle{10}}

\linethickness{0.3mm}
\put(70,87.12){\circle{10}}

\linethickness{0.3mm}
\put(90,87.12){\circle{10}}

\linethickness{0.3mm}
\multiput(25,97.12)(0.24,-0.12){83}{\line(1,0){0.24}}
\linethickness{0.3mm}
\multiput(30,82.12)(0.36,0.12){42}{\line(1,0){0.36}}
\linethickness{0.3mm}
\put(55,87.12){\line(1,0){10}}
\linethickness{0.3mm}
\put(75,87.12){\line(1,0){10}}
\linethickness{0.3mm}
\put(125,162.12){\line(1,0){10}}
\put(125,152.12){\line(0,1){10}}
\put(135,152.12){\line(0,1){10}}
\put(125,152.12){\line(1,0){10}}
\linethickness{0.3mm}
\put(165,152.12){\line(1,0){10}}
\put(165,142.12){\line(0,1){10}}
\put(175,142.12){\line(0,1){10}}
\put(165,142.12){\line(1,0){10}}
\linethickness{0.3mm}
\put(185,152.12){\line(1,0){10}}
\put(185,142.12){\line(0,1){10}}
\put(195,142.12){\line(0,1){10}}
\put(185,142.12){\line(1,0){10}}
\linethickness{0.3mm}
\put(205,152.12){\line(1,0){10}}
\put(205,142.12){\line(0,1){10}}
\put(215,142.12){\line(0,1){10}}
\put(205,142.12){\line(1,0){10}}
\linethickness{0.3mm}
\put(145,142.12){\line(1,0){10}}
\put(145,132.12){\line(0,1){10}}
\put(155,132.12){\line(0,1){10}}
\put(145,132.12){\line(1,0){10}}
\linethickness{0.3mm}
\multiput(135,157.12)(0.36,-0.12){83}{\line(1,0){0.36}}
\put(135,157.12){\vector(-3,1){0.12}}
\linethickness{0.3mm}
\multiput(150,142.12)(0.36,0.12){42}{\line(1,0){0.36}}
\put(165,147.12){\vector(3,1){0.12}}
\linethickness{0.3mm}
\put(175,147.12){\line(1,0){10}}
\put(175,147.12){\vector(-1,0){0.12}}
\linethickness{0.3mm}
\put(195,147.12){\line(1,0){10}}
\put(195,147.12){\vector(-1,0){0.12}}
\linethickness{0.3mm}
\put(125,132.12){\line(1,0){10}}
\put(125,122.12){\line(0,1){10}}
\put(135,122.12){\line(0,1){10}}
\put(125,122.12){\line(1,0){10}}
\linethickness{0.3mm}
\put(165,122.12){\line(1,0){10}}
\put(165,112.12){\line(0,1){10}}
\put(175,112.12){\line(0,1){10}}
\put(165,112.12){\line(1,0){10}}
\linethickness{0.3mm}
\put(185,122.12){\line(1,0){10}}
\put(185,112.12){\line(0,1){10}}
\put(195,112.12){\line(0,1){10}}
\put(185,112.12){\line(1,0){10}}
\linethickness{0.3mm}
\put(205,122.12){\line(1,0){10}}
\put(205,112.12){\line(0,1){10}}
\put(215,112.12){\line(0,1){10}}
\put(205,112.12){\line(1,0){10}}
\linethickness{0.3mm}
\put(145,112.12){\line(1,0){10}}
\put(145,102.12){\line(0,1){10}}
\put(155,102.12){\line(0,1){10}}
\put(145,102.12){\line(1,0){10}}
\linethickness{0.3mm}
\multiput(135,127.12)(0.36,-0.12){83}{\line(1,0){0.36}}
\put(165,117.12){\vector(3,-1){0.12}}
\linethickness{0.3mm}
\multiput(150,112.12)(0.36,0.12){42}{\line(1,0){0.36}}
\put(150,112.12){\vector(-3,-1){0.12}}
\linethickness{0.3mm}
\put(175,117.12){\line(1,0){10}}
\put(175,117.12){\vector(-1,0){0.12}}
\linethickness{0.3mm}
\put(195,117.12){\line(1,0){10}}
\put(195,117.12){\vector(-1,0){0.12}}
\linethickness{0.3mm}
\put(125,102.12){\line(1,0){10}}
\put(125,92.12){\line(0,1){10}}
\put(135,92.12){\line(0,1){10}}
\put(125,92.12){\line(1,0){10}}
\linethickness{0.3mm}
\put(165,92.12){\line(1,0){10}}
\put(165,82.12){\line(0,1){10}}
\put(175,82.12){\line(0,1){10}}
\put(165,82.12){\line(1,0){10}}
\linethickness{0.3mm}
\put(185,92.12){\line(1,0){10}}
\put(185,82.12){\line(0,1){10}}
\put(195,82.12){\line(0,1){10}}
\put(185,82.12){\line(1,0){10}}
\linethickness{0.3mm}
\put(205,92.12){\line(1,0){10}}
\put(205,82.12){\line(0,1){10}}
\put(215,82.12){\line(0,1){10}}
\put(205,82.12){\line(1,0){10}}
\linethickness{0.3mm}
\put(145,82.12){\line(1,0){10}}
\put(145,72.12){\line(0,1){10}}
\put(155,72.12){\line(0,1){10}}
\put(145,72.12){\line(1,0){10}}
\linethickness{0.3mm}
\multiput(135,97.12)(0.36,-0.12){83}{\line(1,0){0.36}}
\put(165,87.12){\vector(3,-1){0.12}}
\linethickness{0.3mm}
\multiput(150,82.12)(0.36,0.12){42}{\line(1,0){0.36}}
\put(165,87.12){\vector(3,1){0.12}}
\linethickness{0.3mm}
\put(175,87.12){\line(1,0){10}}
\put(175,87.12){\vector(-1,0){0.12}}
\linethickness{0.3mm}
\put(195,87.12){\line(1,0){10}}
\put(195,87.12){\vector(-1,0){0.12}}
\linethickness{0.3mm}
\put(125,72.12){\line(1,0){10}}
\put(125,62.12){\line(0,1){10}}
\put(135,62.12){\line(0,1){10}}
\put(125,62.12){\line(1,0){10}}
\linethickness{0.3mm}
\put(165,62.12){\line(1,0){10}}
\put(165,52.12){\line(0,1){10}}
\put(175,52.12){\line(0,1){10}}
\put(165,52.12){\line(1,0){10}}
\linethickness{0.3mm}
\put(185,62.12){\line(1,0){10}}
\put(185,52.12){\line(0,1){10}}
\put(195,52.12){\line(0,1){10}}
\put(185,52.12){\line(1,0){10}}
\linethickness{0.3mm}
\put(205,62.12){\line(1,0){10}}
\put(205,52.12){\line(0,1){10}}
\put(215,52.12){\line(0,1){10}}
\put(205,52.12){\line(1,0){10}}
\linethickness{0.3mm}
\put(145,52.12){\line(1,0){10}}
\put(145,42.12){\line(0,1){10}}
\put(155,42.12){\line(0,1){10}}
\put(145,42.12){\line(1,0){10}}
\linethickness{0.3mm}
\multiput(135,67.12)(0.36,-0.12){83}{\line(1,0){0.36}}
\put(165,57.12){\vector(3,-1){0.12}}
\linethickness{0.3mm}
\multiput(150,52.12)(0.36,0.12){42}{\line(1,0){0.36}}
\put(165,57.12){\vector(3,1){0.12}}
\linethickness{0.3mm}
\put(175,57.12){\line(1,0){10}}
\put(185,57.12){\vector(1,0){0.12}}
\linethickness{0.3mm}
\put(195,57.12){\line(1,0){10}}
\put(195,57.12){\vector(-1,0){0.12}}
\linethickness{0.3mm}
\put(125,42.12){\line(1,0){10}}
\put(125,32.12){\line(0,1){10}}
\put(135,32.12){\line(0,1){10}}
\put(125,32.12){\line(1,0){10}}
\linethickness{0.3mm}
\put(165,32.12){\line(1,0){10}}
\put(165,22.12){\line(0,1){10}}
\put(175,22.12){\line(0,1){10}}
\put(165,22.12){\line(1,0){10}}
\linethickness{0.3mm}
\put(185,32.12){\line(1,0){10}}
\put(185,22.12){\line(0,1){10}}
\put(195,22.12){\line(0,1){10}}
\put(185,22.12){\line(1,0){10}}
\linethickness{0.3mm}
\put(205,32.12){\line(1,0){10}}
\put(205,22.12){\line(0,1){10}}
\put(215,22.12){\line(0,1){10}}
\put(205,22.12){\line(1,0){10}}
\linethickness{0.3mm}
\put(145,22.12){\line(1,0){10}}
\put(145,12.12){\line(0,1){10}}
\put(155,12.12){\line(0,1){10}}
\put(145,12.12){\line(1,0){10}}
\linethickness{0.3mm}
\multiput(135,37.12)(0.36,-0.12){83}{\line(1,0){0.36}}
\put(165,27.12){\vector(3,-1){0.12}}
\linethickness{0.3mm}
\multiput(150,22.12)(0.36,0.12){42}{\line(1,0){0.36}}
\put(165,27.12){\vector(3,1){0.12}}
\linethickness{0.3mm}
\put(175,27.12){\line(1,0){10}}
\put(185,27.12){\vector(1,0){0.12}}
\linethickness{0.3mm}
\put(195,27.12){\line(1,0){10}}
\put(205,27.12){\vector(1,0){0.12}}
\put(20,107.12){\makebox(0,0)[cc]{a}}

\put(30,67.12){\makebox(0,0)[cc]{b}}

\put(50,77.12){\makebox(0,0)[cc]{c}}

\put(70,77.12){\makebox(0,0)[cc]{d}}

\put(90,77.12){\makebox(0,0)[cc]{e}}

\linethickness{0.3mm}
\put(122.88,165){\line(1,0){14.55}}
\put(122.88,28.03){\line(0,1){136.97}}
\put(137.42,28.03){\line(0,1){136.97}}
\put(122.88,28.03){\line(1,0){14.55}}
\linethickness{0.3mm}
\put(202.42,154.85){\line(1,0){14.55}}
\put(202.42,17.88){\line(0,1){136.97}}
\put(216.97,17.88){\line(0,1){136.97}}
\put(202.42,17.88){\line(1,0){14.55}}
\linethickness{0.3mm}
\put(182.88,155){\line(1,0){14.55}}
\put(182.88,18.03){\line(0,1){136.97}}
\put(197.42,18.03){\line(0,1){136.97}}
\put(182.88,18.03){\line(1,0){14.55}}
\linethickness{0.3mm}
\put(162.73,154.85){\line(1,0){14.55}}
\put(162.73,17.88){\line(0,1){136.97}}
\put(177.27,17.88){\line(0,1){136.97}}
\put(162.73,17.88){\line(1,0){14.55}}
\linethickness{0.3mm}
\put(142.88,145){\line(1,0){14.55}}
\put(142.88,8.03){\line(0,1){136.97}}
\put(157.42,8.03){\line(0,1){136.97}}
\put(142.88,8.03){\line(1,0){14.55}}
\put(130,22.12){\makebox(0,0)[cc]{a}}

\put(150,2.12){\makebox(0,0)[cc]{b}}

\put(170,12.12){\makebox(0,0)[cc]{c}}

\put(190,12.12){\makebox(0,0)[cc]{d}}

\put(210,12.12){\makebox(0,0)[cc]{e}}

\end{picture}
\end{centering}
\caption{\label{fig:ExempleBG1}Example of a "destination-based" buffer graph (on the right) on the network of the left.}
\end{figure}

\paragraph{"Destination-based" buffer graph.} In this scheme, we assume that the routing algorithm forwards all packets of Destination $d$ via a directed tree $T_{d}$ rooted in $d$. Each processor $p$ of the network has a buffer $b_{p}(d)$ for each possible Destination $d$ (called the target of $b_{p}(d)$). The buffer graph has $n$ connected components, each of them containing all the buffers which shared their target. The connected component associated to the target $d$ is isomorphic to $T_{d}$. The reader can find an example of a such graph in Figure \ref{fig:ExempleBG1}.

Since each connected component of this graph is a tree, this oriented graph is acyclic. Consequently, \cite{MS78} allows us to define a deadlock-free controller on this graph. Note that this scheme use $n$ buffers per processor. So, we need $n^{2}$ buffers on the whole network.

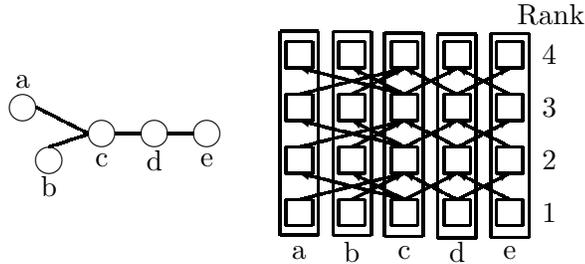
\begin{figure}
\noindent \begin{centering}
\ifx\JPicScale\undefined\def\JPicScale{0.35}\fi
\unitlength \JPicScale mm
\begin{picture}(215.61,90.45)(0,0)
\linethickness{0.3mm}
\put(15,55){\circle{10}}

\linethickness{0.3mm}
\put(45,45){\circle{10}}

\linethickness{0.3mm}
\put(25,35){\circle{10}}

\linethickness{0.3mm}
\put(65,45){\circle{10}}

\linethickness{0.3mm}
\put(85,45){\circle{10}}

\linethickness{0.3mm}
\multiput(20,55)(0.24,-0.12){83}{\line(1,0){0.24}}
\linethickness{0.3mm}
\multiput(25,40)(0.36,0.12){42}{\line(1,0){0.36}}
\linethickness{0.3mm}
\put(50,45){\line(1,0){10}}
\linethickness{0.3mm}
\put(70,45){\line(1,0){10}}
\put(15,65){\makebox(0,0)[cc]{a}}

\put(25,25){\makebox(0,0)[cc]{b}}

\put(45,35){\makebox(0,0)[cc]{c}}

\put(65,35){\makebox(0,0)[cc]{d}}

\put(85,35){\makebox(0,0)[cc]{e}}

\linethickness{0.3mm}
\put(115,20){\line(1,0){10}}
\put(115,10){\line(0,1){10}}
\put(125,10){\line(0,1){10}}
\put(115,10){\line(1,0){10}}
\linethickness{0.3mm}
\put(135,20){\line(1,0){10}}
\put(135,10){\line(0,1){10}}
\put(145,10){\line(0,1){10}}
\put(135,10){\line(1,0){10}}
\linethickness{0.3mm}
\put(155,20){\line(1,0){10}}
\put(155,10){\line(0,1){10}}
\put(165,10){\line(0,1){10}}
\put(155,10){\line(1,0){10}}
\linethickness{0.3mm}
\put(175,20){\line(1,0){10}}
\put(175,10){\line(0,1){10}}
\put(185,10){\line(0,1){10}}
\put(175,10){\line(1,0){10}}
\linethickness{0.3mm}
\put(195,20){\line(1,0){10}}
\put(195,10){\line(0,1){10}}
\put(205,10){\line(0,1){10}}
\put(195,10){\line(1,0){10}}
\put(120,0){\makebox(0,0)[cc]{a}}

\put(140,0){\makebox(0,0)[cc]{b}}

\put(160,0){\makebox(0,0)[cc]{c}}

\put(180,0){\makebox(0,0)[cc]{d}}

\put(200,0){\makebox(0,0)[cc]{e}}

\linethickness{0.3mm}
\multiput(120,20)(0.48,0.12){83}{\line(1,0){0.48}}
\put(160,30){\vector(4,1){0.12}}
\linethickness{0.3mm}
\multiput(140,20)(0.24,0.12){83}{\line(1,0){0.24}}
\put(160,30){\vector(2,1){0.12}}
\linethickness{0.3mm}
\multiput(160,20)(0.24,0.12){83}{\line(1,0){0.24}}
\put(180,30){\vector(2,1){0.12}}
\linethickness{0.3mm}
\multiput(140,30)(0.24,-0.12){83}{\line(1,0){0.24}}
\put(140,30){\vector(-2,1){0.12}}
\linethickness{0.3mm}
\multiput(120,30)(0.48,-0.12){83}{\line(1,0){0.48}}
\put(120,30){\vector(-4,1){0.12}}
\linethickness{0.3mm}
\multiput(180,20)(0.24,0.12){83}{\line(1,0){0.24}}
\put(200,30){\vector(2,1){0.12}}
\linethickness{0.3mm}
\multiput(160,30)(0.24,-0.12){83}{\line(1,0){0.24}}
\put(160,30){\vector(-2,1){0.12}}
\linethickness{0.3mm}
\multiput(180,30)(0.24,-0.12){83}{\line(1,0){0.24}}
\put(180,30){\vector(-2,1){0.12}}
\linethickness{0.3mm}
\put(115,40){\line(1,0){10}}
\put(115,30){\line(0,1){10}}
\put(125,30){\line(0,1){10}}
\put(115,30){\line(1,0){10}}
\linethickness{0.3mm}
\put(135,40){\line(1,0){10}}
\put(135,30){\line(0,1){10}}
\put(145,30){\line(0,1){10}}
\put(135,30){\line(1,0){10}}
\linethickness{0.3mm}
\put(155,40){\line(1,0){10}}
\put(155,30){\line(0,1){10}}
\put(165,30){\line(0,1){10}}
\put(155,30){\line(1,0){10}}
\linethickness{0.3mm}
\put(175,40){\line(1,0){10}}
\put(175,30){\line(0,1){10}}
\put(185,30){\line(0,1){10}}
\put(175,30){\line(1,0){10}}
\linethickness{0.3mm}
\put(195,40){\line(1,0){10}}
\put(195,30){\line(0,1){10}}
\put(205,30){\line(0,1){10}}
\put(195,30){\line(1,0){10}}
\linethickness{0.3mm}
\multiput(120,40)(0.48,0.12){83}{\line(1,0){0.48}}
\put(160,50){\vector(4,1){0.12}}
\linethickness{0.3mm}
\multiput(140,40)(0.24,0.12){83}{\line(1,0){0.24}}
\put(160,50){\vector(2,1){0.12}}
\linethickness{0.3mm}
\multiput(160,40)(0.24,0.12){83}{\line(1,0){0.24}}
\put(180,50){\vector(2,1){0.12}}
\linethickness{0.3mm}
\multiput(140,50)(0.24,-0.12){83}{\line(1,0){0.24}}
\put(140,50){\vector(-2,1){0.12}}
\linethickness{0.3mm}
\multiput(120,50)(0.48,-0.12){83}{\line(1,0){0.48}}
\put(120,50){\vector(-4,1){0.12}}
\linethickness{0.3mm}
\multiput(180,40)(0.24,0.12){83}{\line(1,0){0.24}}
\put(200,50){\vector(2,1){0.12}}
\linethickness{0.3mm}
\multiput(160,50)(0.24,-0.12){83}{\line(1,0){0.24}}
\put(160,50){\vector(-2,1){0.12}}
\linethickness{0.3mm}
\multiput(180,50)(0.24,-0.12){83}{\line(1,0){0.24}}
\put(180,50){\vector(-2,1){0.12}}
\linethickness{0.3mm}
\put(115,60){\line(1,0){10}}
\put(115,50){\line(0,1){10}}
\put(125,50){\line(0,1){10}}
\put(115,50){\line(1,0){10}}
\linethickness{0.3mm}
\put(135,60){\line(1,0){10}}
\put(135,50){\line(0,1){10}}
\put(145,50){\line(0,1){10}}
\put(135,50){\line(1,0){10}}
\linethickness{0.3mm}
\put(155,60){\line(1,0){10}}
\put(155,50){\line(0,1){10}}
\put(165,50){\line(0,1){10}}
\put(155,50){\line(1,0){10}}
\linethickness{0.3mm}
\put(175,60){\line(1,0){10}}
\put(175,50){\line(0,1){10}}
\put(185,50){\line(0,1){10}}
\put(175,50){\line(1,0){10}}
\linethickness{0.3mm}
\put(195,60){\line(1,0){10}}
\put(195,50){\line(0,1){10}}
\put(205,50){\line(0,1){10}}
\put(195,50){\line(1,0){10}}
\linethickness{0.3mm}
\multiput(120,60)(0.48,0.12){83}{\line(1,0){0.48}}
\put(160,70){\vector(4,1){0.12}}
\linethickness{0.3mm}
\multiput(140,60)(0.24,0.12){83}{\line(1,0){0.24}}
\put(160,70){\vector(2,1){0.12}}
\linethickness{0.3mm}
\multiput(160,60)(0.24,0.12){83}{\line(1,0){0.24}}
\put(180,70){\vector(2,1){0.12}}
\linethickness{0.3mm}
\multiput(140,70)(0.24,-0.12){83}{\line(1,0){0.24}}
\put(140,70){\vector(-2,1){0.12}}
\linethickness{0.3mm}
\multiput(120,70)(0.48,-0.12){83}{\line(1,0){0.48}}
\put(120,70){\vector(-4,1){0.12}}
\linethickness{0.3mm}
\multiput(180,60)(0.24,0.12){83}{\line(1,0){0.24}}
\put(200,70){\vector(2,1){0.12}}
\linethickness{0.3mm}
\multiput(160,70)(0.24,-0.12){83}{\line(1,0){0.24}}
\put(160,70){\vector(-2,1){0.12}}
\linethickness{0.3mm}
\multiput(180,70)(0.24,-0.12){83}{\line(1,0){0.24}}
\put(180,70){\vector(-2,1){0.12}}
\linethickness{0.3mm}
\put(115,80){\line(1,0){10}}
\put(115,70){\line(0,1){10}}
\put(125,70){\line(0,1){10}}
\put(115,70){\line(1,0){10}}
\linethickness{0.3mm}
\put(135,80){\line(1,0){10}}
\put(135,70){\line(0,1){10}}
\put(145,70){\line(0,1){10}}
\put(135,70){\line(1,0){10}}
\linethickness{0.3mm}
\put(155,80){\line(1,0){10}}
\put(155,70){\line(0,1){10}}
\put(165,70){\line(0,1){10}}
\put(155,70){\line(1,0){10}}
\linethickness{0.3mm}
\put(175,80){\line(1,0){10}}
\put(175,70){\line(0,1){10}}
\put(185,70){\line(0,1){10}}
\put(175,70){\line(1,0){10}}
\linethickness{0.3mm}
\put(195,80){\line(1,0){10}}
\put(195,70){\line(0,1){10}}
\put(205,70){\line(0,1){10}}
\put(195,70){\line(1,0){10}}
\put(215.61,90.45){\makebox(0,0)[cc]{Rank}}

\put(215,15){\makebox(0,0)[cc]{1}}

\put(215,35){\makebox(0,0)[cc]{2}}

\put(215,55){\makebox(0,0)[cc]{3}}

\put(215,75){\makebox(0,0)[cc]{4}}

\linethickness{0.3mm}
\put(112.73,83.94){\line(1,0){14.55}}
\put(112.73,6.52){\line(0,1){77.42}}
\put(127.27,6.52){\line(0,1){77.42}}
\put(112.73,6.52){\line(1,0){14.55}}
\linethickness{0.3mm}
\put(132.88,83.79){\line(1,0){14.55}}
\put(132.88,6.36){\line(0,1){77.42}}
\put(147.42,6.36){\line(0,1){77.42}}
\put(132.88,6.36){\line(1,0){14.55}}
\linethickness{0.3mm}
\put(152.42,83.33){\line(1,0){14.55}}
\put(152.42,5.91){\line(0,1){77.42}}
\put(166.97,5.91){\line(0,1){77.42}}
\put(152.42,5.91){\line(1,0){14.55}}
\linethickness{0.3mm}
\put(172.58,82.88){\line(1,0){14.55}}
\put(172.58,5.45){\line(0,1){77.42}}
\put(187.12,5.45){\line(0,1){77.42}}
\put(172.58,5.45){\line(1,0){14.55}}
\linethickness{0.3mm}
\put(192.73,83.03){\line(1,0){14.55}}
\put(192.73,5.61){\line(0,1){77.42}}
\put(207.27,5.61){\line(0,1){77.42}}
\put(192.73,5.61){\line(1,0){14.55}}
\end{picture}
\end{centering}
\caption{\label{fig:ExempleBG2}Example of a "distance-based" buffer graph (on the right) on the network of the left.}
\end{figure}

\paragraph{"Distance-based" buffer graph.} In this scheme, each processor have $D+1$ buffers ranked from $1$ to $D+1$ (remind that $D$ is the diameter of the network). New messages are always generated in the buffer of rank $1$ of the sending processor. When a message occupying a buffer of rank $i$ is forwarded to a neighbor $q$, it is always copied in the buffer of rank $i+1$ of $q$. We need $D+1$ buffers per processor since, in the worst case, there are $D$ forwarding of a message between its generation and its consumption. The reader can find an example of such a graph in Figure \ref{fig:ExempleBG2}.

Since messages always "come upstairs" the buffer rank, this oriented graph is acyclic. Consequently, \cite{MS78} allows us to define a deadlock-free controller on this graph. Note that this scheme use $D+1$ buffers per processor. So, we need $n(D+1)$ buffers on the whole network.

\section{First protocol}\label{sec:protocolN}

\subsection{Informal description}

The main idea of this section is to adapt the "destination-based" scheme (see Section \ref{sec:survey}) in order to tolerate the corruption of routing tables in the initial configuration. To perform this goal, we assume the existence of a self-stabilizing silent (\emph{i.e.} no actions are enabled after convergence) algorithm $\mathcal{A}$ to compute routing tables which runs simultaneously to our message forwarding protocol. Moreover, we assume that $\mathcal{A}$ has priority over our protocol (\emph{i.e.} a processor which has enabled actions for both algorithms always chooses the action of $\mathcal{A}$). This guarantees us that routing tables are correct and constant in a finite time. To simplify the presentation, we assume that $\mathcal{A}$ induces only minimal paths in number of edges. We assume that our protocol can have access to the routing table via a function, called $nextHop_{p}(d)$. This function returns the identity of the neighbor of $p$ to which $p$ must forward messages of Destination $d$. 

\begin{figure}
\noindent \begin{centering}
\ifx\JPicScale\undefined\def\JPicScale{0.35}\fi
\unitlength \JPicScale mm
\begin{picture}(217.27,85)(0,0)
\linethickness{0.3mm}
\put(20,52.12){\circle{10}}

\linethickness{0.3mm}
\put(50,42.12){\circle{10}}

\linethickness{0.3mm}
\put(30,32.12){\circle{10}}

\linethickness{0.3mm}
\put(70,42.12){\circle{10}}

\linethickness{0.3mm}
\put(90,42.12){\circle{10}}

\linethickness{0.3mm}
\multiput(25,52.12)(0.24,-0.12){83}{\line(1,0){0.24}}
\linethickness{0.3mm}
\multiput(30,37.12)(0.36,0.12){42}{\line(1,0){0.36}}
\linethickness{0.3mm}
\put(55,42.12){\line(1,0){10}}
\linethickness{0.3mm}
\put(75,42.12){\line(1,0){10}}
\put(20,62.12){\makebox(0,0)[cc]{a}}

\put(30,22.12){\makebox(0,0)[cc]{b}}

\put(50,32.12){\makebox(0,0)[cc]{c}}

\put(70,32.12){\makebox(0,0)[cc]{d}}

\put(90,32.12){\makebox(0,0)[cc]{e}}

\linethickness{0.3mm}
\put(125,60){\line(1,0){10}}
\put(125,50){\line(0,1){10}}
\put(135,50){\line(0,1){10}}
\put(125,50){\line(1,0){10}}
\linethickness{0.3mm}
\put(165,60){\line(1,0){10}}
\put(165,50){\line(0,1){10}}
\put(175,50){\line(0,1){10}}
\put(165,50){\line(1,0){10}}
\linethickness{0.3mm}
\put(185,60){\line(1,0){10}}
\put(185,50){\line(0,1){10}}
\put(195,50){\line(0,1){10}}
\put(185,50){\line(1,0){10}}
\linethickness{0.3mm}
\put(205,60){\line(1,0){10}}
\put(205,50){\line(0,1){10}}
\put(215,50){\line(0,1){10}}
\put(205,50){\line(1,0){10}}
\linethickness{0.3mm}
\put(145,40){\line(1,0){10}}
\put(145,30){\line(0,1){10}}
\put(155,30){\line(0,1){10}}
\put(145,30){\line(1,0){10}}
\linethickness{0.3mm}
\put(122.73,85){\line(1,0){14.54}}
\put(122.73,45){\line(0,1){40}}
\put(137.27,45){\line(0,1){40}}
\put(122.73,45){\line(1,0){14.54}}
\linethickness{0.3mm}
\put(202.72,65){\line(1,0){14.55}}
\put(202.72,25){\line(0,1){40}}
\put(217.27,25){\line(0,1){40}}
\put(202.72,25){\line(1,0){14.55}}
\linethickness{0.3mm}
\put(182.73,65){\line(1,0){14.54}}
\put(182.73,25){\line(0,1){40}}
\put(197.27,25){\line(0,1){40}}
\put(182.73,25){\line(1,0){14.54}}
\linethickness{0.3mm}
\put(162.73,65){\line(1,0){14.54}}
\put(162.73,25){\line(0,1){40}}
\put(177.27,25){\line(0,1){40}}
\put(162.73,25){\line(1,0){14.54}}
\linethickness{0.3mm}
\put(142.73,45){\line(1,0){14.54}}
\put(142.73,5){\line(0,1){40}}
\put(157.27,5){\line(0,1){40}}
\put(142.73,5){\line(1,0){14.54}}
\linethickness{0.3mm}
\put(125,80){\line(1,0){10}}
\put(125,70){\line(0,1){10}}
\put(135,70){\line(0,1){10}}
\put(125,70){\line(1,0){10}}
\linethickness{0.3mm}
\put(145,20){\line(1,0){10}}
\put(145,10){\line(0,1){10}}
\put(155,10){\line(0,1){10}}
\put(145,10){\line(1,0){10}}
\linethickness{0.3mm}
\put(165,40){\line(1,0){10}}
\put(165,30){\line(0,1){10}}
\put(175,30){\line(0,1){10}}
\put(165,30){\line(1,0){10}}
\linethickness{0.3mm}
\put(185,40){\line(1,0){10}}
\put(185,30){\line(0,1){10}}
\put(195,30){\line(0,1){10}}
\put(185,30){\line(1,0){10}}
\linethickness{0.3mm}
\put(205,40){\line(1,0){10}}
\put(205,30){\line(0,1){10}}
\put(215,30){\line(0,1){10}}
\put(205,30){\line(1,0){10}}
\put(130,35){\makebox(0,0)[cc]{a}}

\put(150,0){\makebox(0,0)[cc]{b}}

\put(170,20){\makebox(0,0)[cc]{c}}

\put(190,20){\makebox(0,0)[cc]{d}}

\put(210,20){\makebox(0,0)[cc]{e}}

\linethickness{0.3mm}
\put(130,60){\line(0,1){10}}
\put(130,60){\vector(0,-1){0.12}}
\linethickness{0.3mm}
\put(170,40){\line(0,1){10}}
\put(170,40){\vector(0,-1){0.12}}
\linethickness{0.3mm}
\put(190,40){\line(0,1){10}}
\put(190,40){\vector(0,-1){0.12}}
\linethickness{0.3mm}
\put(210,40){\line(0,1){10}}
\put(210,40){\vector(0,-1){0.12}}
\linethickness{0.3mm}
\put(150,20){\line(0,1){10}}
\put(150,20){\vector(0,-1){0.12}}
\linethickness{0.3mm}
\put(135,55){\line(1,0){30}}
\put(165,55){\vector(1,0){0.12}}
\linethickness{0.3mm}
\multiput(175,55)(0.12,-0.24){83}{\line(0,-1){0.24}}
\put(175,55){\vector(-1,2){0.12}}
\linethickness{0.3mm}
\multiput(195,55)(0.12,-0.24){83}{\line(0,-1){0.24}}
\put(195,55){\vector(-1,2){0.12}}
\linethickness{0.3mm}
\put(155,35){\line(1,0){10}}
\put(155,35){\vector(-1,0){0.12}}
\end{picture}
\end{centering}
\caption{\label{fig:ExempleBG3}Example of our buffer graph (on the right) for Destination $b$ on the network (on the left).}
\end{figure}
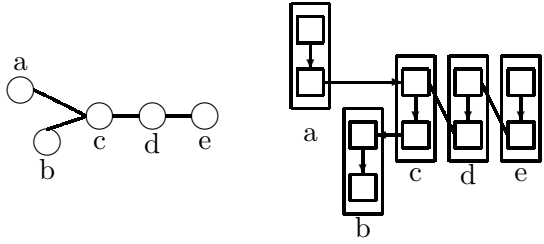

We now describe our buffer graph adapted from the "destination-based" one. Our buffer graph is composed of $n$ connected components, each associated to a destination $d$ and based on the oriented tree $T_{d}$ (remind that $T_{d}$ is the tree induced by routing table for Destination $d$). Consequently, we can present only one connected component, associated to a destination noted $d$ (others are similar). We use two buffers per processor for Destination $d$. The first one, noted $bufR_{p}(d)$ (for processor $p$), is reserved to the reception of messages whereas the second one, noted $bufE_{p}(d)$, is used to emit messages (see Figure \ref{fig:ExempleBG3}). This scheme allows us to control the advance of messages. Indeed, we allow a message to be forwarded from $bufR_{p}(d)$ to $bufE_{p}(d)$ if and only if the message is only present in $bufR_{p}(d)$ and we erase it simultaneously. In this way, we can control the consequences of routing tables moves on messages (duplication or merge which can involve message losses).

To avoid livelocks, we use a fair scheme of selection of processors allowed to forward or to emit a message for each reception buffer. We can manage this fairness by a queue of requesting processors. Finally, we use a specific flag to prevent message losses. It is composed of the identity of the last processor cross over by the message and a $color$ which is dynamically given to the message when it reaches an emission buffer. In order to distinguish a such incoming message of these contained in reception buffers of neighbors of the considered processor, we give to this incoming message a $color$ which is not carried by a such message. It is why a message is considered as a triplet $(m,p,c)$ in our algorithm where $m$ is the useful information of the message, $p$ is the identity of the last processor crossed over by the message, and $c$ is a color (a natural integer between $0$ and $\Delta$).

We must manage a communication between our algorithm and processors in order to know when a processor have a message to send. We have chosen to create a Boolean shared variable $request_{p}$ (for any processor $p$). Processor $p$ can set it at $true$ when it is at $false$ and when $p$ has a message to send. Otherwise, $p$ must wait that our algorithm sets the shared variable to $false$ (that is done when a message is generated).

\begin{figure}
\noindent \begin{centering}
\includegraphics[scale=0.29]{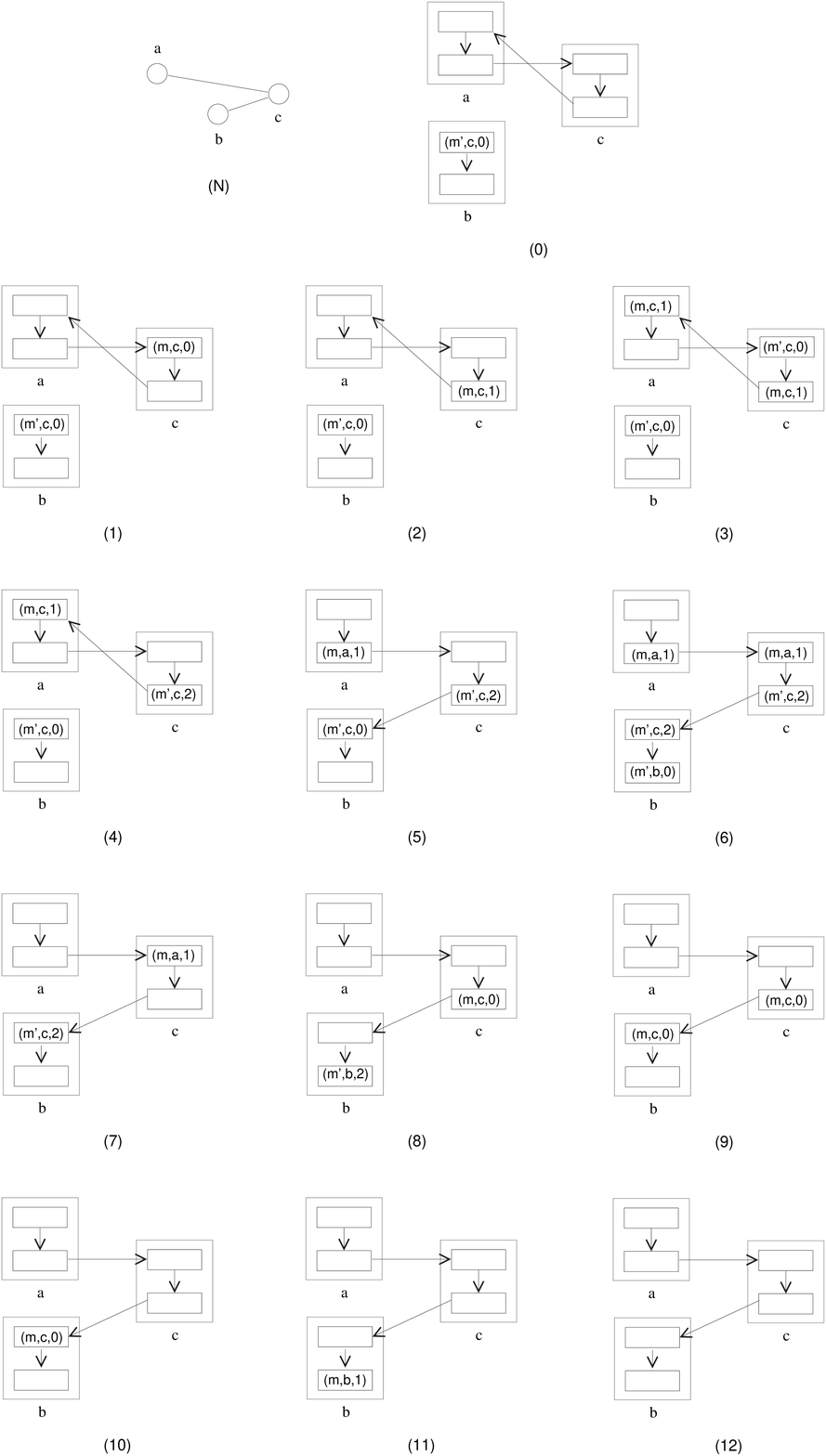}
\par\end{centering}
\caption{\label{fig:ExempleExecution}An example of execution of our first algorithm.}
\end{figure}

The reader can find a complete example of the execution of our algorithm in Figure \ref{fig:ExempleExecution}. Diagram $(N)$ shows the network and diagram $(0)$ shows the initial configuration for the connected component associated to $b$ of the buffer graph. We observe that $\Delta=3$, so we need 4 different values for the variable $color$, we have chosen to represent them by a natural integer in $\{0,1,2,3\}$.  Remark that routing tables are incorrect (in particular there exists a cycle involving buffers of $a$ and $c$) and that there exists an invalid message $m'$ in the reception buffer of $b$ (its $color$ is $0$). Then, Processor $c$ emits a message $m$ (its $color$ is $0$) in the reception buffer of $c$ to obtain configuration $(1)$. When the message $m$ is forwarded to the emission buffer of $c$, we associate it the $color$ $1$ (since $0$ is forbidden, see configuration $(2)$). During the next step, message $m$ is forwarded to the reception buffer of $a$ (remark that it keeps its $color$) and $c$ emits (in its reception buffer) a new message $m'$ which has the same useful information as the invalid message present on $b$. So, we obtain configuration $(3)$. Message $m$ can now be erased from the emission buffer of $c$ and $m'$ can be forwarded into this buffer (we associate it the $color$ $2$). These two steps lead to configuration $(4)$. Assume that routing tables are repaired during the next step. Simultaneously, processor $a$ is allowed to forward $m$ into its emission buffer. We obtain configuration $(5)$. Remark that the use of $color$ forbids the merge between the two messages which have $m'$ for useful information. Then, the system is able to deliver these three messages by the repetition of moves that we have described: 
\begin{itemize}
\item forwarding from reception buffer to emission buffer of the same processor.
\item forwarding from emission buffer to reception buffer of two processors.
\item erasing from emission buffer or delivering.
\end{itemize}
The sequence of configuration $(6)$ to $(12)$ shows an example of the end of our execution.

\subsection{Algorithm}

We now present formally our protocol in Algorithm \ref{algo:AN}. We call it \AN for $\mathcal{S}$nap-$\mathcal{S}$tabilizing $\mathcal{M}$essage $\mathcal{F}$orwarding $\mathcal{P}$rotocol 1. In order to simplify the presentation, we write the algorithm for Destination $d$ only. Obviously, each destination of the network needs a similar algorithm. Moreover, we assume that all these algorithms run simultaneously (as they are mutually independent, this assumption has no effect on the provided proof).

\begin{algorithm}
\caption{\label{algo:AN}(\AN): Message forwarding protocol for Processor $p$ with Destination $d$.}
\small{
\begin{description}
\item [Data:] ~\\
- $n$: natural integer equals to the number of processors of the network.\\
- $I=\{0,...,n-1\}$: set of processor identities of the network.\\
- $N_{p}$: set of neighbors of $p$.\\
- $\Delta$: natural integer equals to the maximal degree of the network.
\item [Message:] ~\\
- $(m,q,c)$ with $m$ useful information of the message, $q\in N_{p}\cup\{p\}$ identity of the last processor crossed over by the message, and $c\in\{0,...,\Delta\}$ a color. The message destination is the buffer index.
\item [Variables:] ~\\
- $bufR_{p}(d)$, $bufE_{p}(d)$: buffers which can contain a message or be empty (denoted by $\varepsilon$).
\item [Input/Output:] ~\\
- $request_{p}$: Boolean. The higher layer can set it to $true$ when its value is $false$ and when there is a waiting message. We consider that this waiting is blocking.
\item [Macros:] ~\\
- $nextMessage_{p}$: gives the message waiting in the higher layer.\\
- $nextDestination_{p}$: gives the destination of $nextMessage_{p}$ if it exists, $null$ otherwise.
\item [Procedures:] ~\\
- $nextHop_{p}(d)$: neighbor of $p$ given by the routing algorithm for Destination $d$.\\
- $choice_{p}(d)$: fairly chooses one of the processors which can forward or generate a message in $bufR_{p}(d)$, \emph{i.e.} $choice_{p}(d)$ satisfies predicate $(choice_{p}(d)\in N_{p}\wedge bufE_{choice_{p}(d)}(d)=(m,q,c)\wedge$ $nextHop_{choice_{p}(d)}(d)=p)\vee (choice_{p}(d)=p\wedge request_{p})$. We can manage this fairness with a queue of length $\Delta+1$ of processors which satisfies the predicate.\\
- $deliver{}_{p}(m)$: delivers the message $m$ to the higher layer of $p$.\\
- $color_{p}(d)$: gives a natural integer $c$ between $0$ and $\Delta$ such as $\forall q\in N_{p}$, $bufR_{q}(d)$ does not contain a message with $c$ as color.
\item [Rules:] ~\\
\textbf{/{*}} Rule for the generation of a message \textbf{{*}/}\\
$\boldsymbol{(R_{1})\,}::\, request_{p}\wedge(nextDestination_{p}=d)\wedge(bufR_{p}(d)=\varepsilon)\wedge(choice_{p}(d)=p)\longrightarrow bufR_{p}(d):=(nextMessage_{p},p,0);request_{p}:=false$\\
\textbf{/{*}} Rule for the internal forwarding of a message \textbf{{*}/}\\
$\boldsymbol{(R_{2})}\,::\,(bufE_{p}(d)=\varepsilon)\wedge(bufR_{p}(d)=(m,q,c))\wedge((q=p)\vee(bufE_{q}(d)\neq(m,q',c)))\longrightarrow bufE_{p}(d):=(m,p,color_{p}(d));bufR_{p}(d):=\varepsilon$\\
\textbf{/{*}} Rule for the forwarding of a message \textbf{{*}/}\\
$\boldsymbol{(R_{3})}\,::\,(bufR_{p}(d)=\varepsilon)\wedge(choice_{p}(d)=s)\wedge(s\neq p)\wedge(bufE_{s}(d)=(m,q,c))\longrightarrow bufR_{p}(d):=(m,s,c)^{1}$\\
\textbf{/{*}} Rule for the erasing of a message after its forwarding
\textbf{{*}/}\\
$\boldsymbol{(R_{4})}\,::\,(bufE_{p}(d)=(m,q,c))\wedge(p\neq d)\wedge(bufR_{nextHop_{p}(d)}(d)=(m,p,c))\wedge(\forall r\in N_{p}\backslash\{nextHop_{p}(d)\},\, bufR_{r}(d)\neq(m,p,c))\longrightarrow bufE_{p}(d):=\varepsilon$\\
\textbf{/{*}} Rule for the erasing of a message after its duplication
\textbf{{*}/}\\
$\boldsymbol{(R_{5})}\,::\,(bufR_{p}(d)=(m,q,c))\wedge(bufE_{q}(d)=(m,q',c))\wedge(nextHop_{q}(d)\neq p)\longrightarrow bufR_{p}(d):=\varepsilon$\\
\textbf{/{*}} Rule for the consumption of a message \textbf{{*}/}\\
$\boldsymbol{(R_{6})}\,::\,(bufE_{p}(p)=(m,q,c))\longrightarrow deliver_{p}(m);bufE_{p}(p):=\varepsilon$
\end{description}

$^{1}$ The fact that $q$ may be different of $s$ implies that the message was in the system at the initial configuration. We could locally delete this message but this does not improve the performance of \AN.}
\end{algorithm}

\subsection{Proof of correctness}

In order to simplify the proof, we introduce a second specification of the problem. This specification allows message duplications.

\begin{specification} [\textbf{$\mathcal{SP}'$}] \label{specif:spe1}
Specification of message forwarding problem allowing duplication.
\begin{itemize}
\item Any message can be generated in a finite time.
\item Any valid message is deliver to its destination in a finite time.
\end{itemize}
\end{specification}

In this section, we prove that \AN is a snap-stabilizing message forwarding protocol for specification $\mathcal{SP}$.  For that, we are going to prove successively that:
\begin{enumerate}
\item \AN is a snap-stabilizing message forwarding protocol for specification $\mathcal{SP}'$ if routing tables are correct in the initial configuration (Lemmas \ref{lem:avanceN}, \ref{lem:depotN}, \ref{lem:transportN} and Proposition \ref{prop:snapTRN}). 
\item \AN is a self-stabilizing message forwarding protocol for specification $\mathcal{SP}'$ even if routing tables are corrupted in the initial configuration (Proposition \ref{prop:selfN}). 
\item \AN is a snap-stabilizing message forwarding protocol for specification $\mathcal{SP}$ even if routing tables are corrupted in the initial configuration (Lemmas \ref{lem:perteN}, \ref{lem:duplicationN} and Theorem \ref{th:snapN}).
\end{enumerate}

In this proof, we consider that the notion of message is different from the notion of useful information. This implies that two messages with the same useful information generated by the same processor are considered as two different messages. We must prove that the algorithm does not lose one of them thanks to the use of the flag. Let $\gamma$ be a configuration of the network. We say that a message $m$ is existing in $\gamma$ if at least one buffer contains $m$ in $\gamma$. We say that $m$ is existing on $p$ in $\gamma$ if at least one buffer of $p$ contains $m$ in $\gamma$.

\begin{definition} [Caterpillar of a message $m$] \label{def:caterpillarN}
Let $m$ be a message of Destination $d$ existing on a processor $p$ in a configuration $\gamma$. We define a caterpillar associated to $m$ as the longest sequence of buffers that satisfies one of the three definitions below:
\begin{enumerate}
\item Caterpillar of type 1: $(bufR_{p}(d)=(m,q,c))\wedge((bufE_{q}(d)\neq(m,q',c))\vee(q=p))$. 
\item Caterpillar of type 2: $(bufE_{p}(d)=(m,q,c))\wedge(bufR_{nextHop_{p}(d)}(d)\neq(m,p,c))$. 
\item Caterpillar of type 3: $(bufE_{p}(d)=(m,q',c))\wedge\exists q\in N_{p},\;(bufR_{q}(d)=(m,p,c))$. 
\end{enumerate}
\end{definition}

\begin{figure}
\begin{centering}
\includegraphics[scale=0.35]{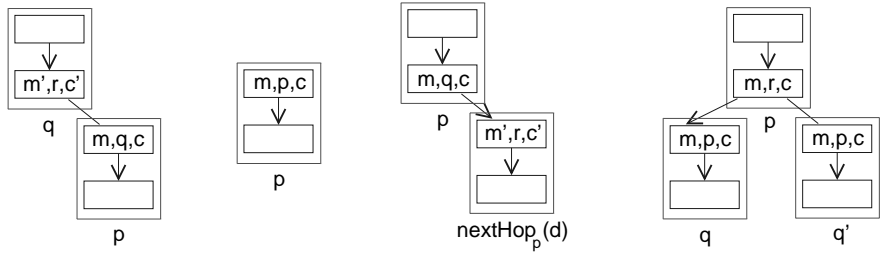}
\par\end{centering}
\caption{\label{fig:CaterpillarN}Examples of caterpillar associated to $m$ on $p$ (from left to right: two of type 1, one of type 2 and one of type 3).}
\end{figure}

The reader can find in Figure \ref{fig:CaterpillarN} an example for each type of caterpillar. Remark that an emission buffer can belong to several caterpillars of type 3.

\begin{lemma} \label{lem:avanceN}
Let $\gamma$ be a configuration in which routing tables are correct. Let $m$ be a message existing on $p$ in $\gamma$. Under a weakly fair daemon, the execution of \AN products in a finite time one of the following effects for any caterpillar of type 1 associated to $m$:
\begin{itemize}
\item $m$ is delivered to its destination.
\item the caterpillar disappeared on $p$ and there exists a caterpillar of type 1 associated to the same message on $nextHop_{p}(d)$. 
\end{itemize}
\end{lemma}

\begin{proof}
Let $\gamma$ be a configuration in which routing tables are correct. Let $m$ (of Destination $d$) be a message existing in $\gamma$. Let $C=bufR_{p}(d)$ be a caterpillar of type 1 associated to $m$. Denote by $\delta$ the distance between $p$ and $d$ ($\delta=dist(p,d)$). We are going to prove the result by induction on $\delta$. We define the following predicate:

$\boldsymbol{(P_{\delta})}$: if $C=bufR_{p}(d)$ is a caterpillar of type 1 associated to $m$ such that $dist(p,d)=\delta$, then, under a weakly fair daemon, the execution of \AN products one of the following effect in a finite time:

\begin{itemize}
\item $m$ is delivered to $d$. 
\item $C$ disappeared on $p$ and there exists a caterpillar of type 1 associated to the same message on $nextHop_{p}(d)$. 
\end{itemize}

\begin{description}
\item [Initialization:] We are going to prove that $\boldsymbol{(P_{0})}$ is true.\\
Let $C=bufR_{p}(d)$ be a caterpillar of type 1 associated to $m$ such that $dist(p,d)=0$. This implies that $p=d$. Let be $bufR_{p}(d)=(m,q,id)$. We must distinguish two cases :

\begin{description}
\item [Case 1:] $bufE_{p}(d)\neq \varepsilon$.\\
The rule $\boldsymbol{(R_{6})}$ is enabled for the processor $p$. We can observe that this rule can not be neutralized. Since we assumed a weakly fair daemon, we obtain that $p$ executes $\boldsymbol{(R_{6})}$ in a finite time. We can then consider the case 2 since this rule erases the content of $bufE_{p}(d)$.
\item [Case 2:] $bufE_{p}(d)=\varepsilon$.\\
By the definition of a caterpillar of type 1, $\boldsymbol{(R_{2})}$ is enabled for $p$. This rule can be neutralized if and only if $bufE_{q}(d)$ is occupied by $(m,q',id)$. This is impossible by the construction of $color_{q}(d)$. Since we assume a weakly fair daemon, we obtain that $p$ executes $\boldsymbol{(R_{2})}$ in a finite time. $C$ disappears and a new caterpillar of type 2 appears in $bufE_{p}(d)$. By the same reasoning of the case 1, we can say that $p$ executes $\boldsymbol{(R_{6})}$ in a finite time. This implies that $m$ is delivered to $d$.
\end{description}

We proved that $\boldsymbol{(P_{0})}$ is true.

\item [Induction:] Let $\delta\geq1$. We assume that $\boldsymbol{(P_{\delta-1})}$ is true. We are going to prove that then $\boldsymbol{(P_{\delta})}$ is true.\\
Let $C=bufR_{p}(d)$ be a caterpillar of type 1 associated to $m$ such that $dist(p,d)=\delta$. Let be $bufR_{p}(d)=(m,q,id)$. We must distinguish two cases:

\begin{description}
\item [Case 1:] $bufE_{p}(d)\neq \varepsilon$.\\
Let be $r=nextHop_{p}(d)$.

\begin{description}
\item [Case 1.1:] $bufE_{p}(d)$ is occupied by a caterpillar $C'$ of type 2.\\
By the definition of a caterpillar of type 2, either $\boldsymbol{(R_{3})}$ or $\boldsymbol{(R_{1})}$ is enabled on $r$ if and only if $bufR_{r}(d)=\varepsilon$.

\begin{description}
\item [Case 1.1.a:] If $bufR_{r}(d)=\varepsilon$, then $r$ executes $\boldsymbol{(R_{3})}$ or $\boldsymbol{(R_{1})}$ (since we assumed a weakly fair daemon and these rules cannot be neutralized). The result of this execution depends on the value of $choice_{r}(d)$:

\begin{itemize}
\item If $choice_{r}(d)=p$, then $C'$ becomes a caterpillar of type 3. We are now in the case 1.2. 
\item If $choice_{r}(d)\neq p$, then a message $(m',$ $choice_{r}(d),id')$ is forwarded in $bufR_{r}(d)$. So, $C'$ remains a caterpillar of type 2 and we are in the case 1.1.b. It is important to remark that the fairness of $choice_{r}(d)$ guarantees us that this case cannot appear infinitely. 
\end{itemize}

\item [Case 1.1.b:] If $bufR_{r}(d)=(m',q',id')$, then we can distinguish two cases:

\begin{itemize}
\item If $bufR_{r}(d)$ belongs to at least one caterpillar of type 3, we can apply the reasoning of the case 1.2 to $bufE_{q'}(d)$ and conclude that $bufR_{r}(d)$ belongs to a caterpillar of type 1 in a finite time. 
\item If $bufR_{r}(d)$ belongs to a caterpillar of type 1, we can say that $bufR_{r}(d)$ becomes empty in a finite time by application of $\boldsymbol{(P_{\delta-1})}$ ($dist(r,d)=dist(p,d)-1=\delta-1$ since routing tables are correct). Then, we are on the case 1.1.a. 
\end{itemize}

\end{description}

We can conclude that $bufE_{p}(d)$ belongs to a caterpillar of type 3 associated to $m$ in a finite time. So, we are on the case 1.2.

\item [Case 1.2:] $bufE_{p}(d)$ belongs to at least one caterpillar of type 3.

\begin{description}
\item [Case 1.2.a:] $bufE_{p}(d)$ belongs to at least two caterpillars of type 3.\\
This implies that there exists $x\in N_{p}\backslash\{r\}$, $bufR_{x}(d)=(m,p,id)$. The processor $x$ is enabled by $\boldsymbol{(R_{5})}$ infinitely (since routing tables are correct and $p$ cannot erase $bufE_{p}(d)$ by the construction of $\boldsymbol{(R_{4})}$). Since we assumed a weakly fair daemon, $(m,p,id$) is erased from $bufR_{x}(d)$ in a finite time. We can repeat this reasoning until $bufE_{p}(d)$ belongs to only one caterpillar of type 3 since the construction of $\boldsymbol{(R_{3})}$ guarantees us that it is impossible to create a new caterpillar of type $3$ involving $bufE_{p}(d)$. So, we are on the case 1.2.b.
\item [Case 1.2.b:] $bufE_{p}(d)$ belongs to only one caterpillar of type 3.\\
By the definition of a caterpillar of type 3, we can say that $p$ is enabled for $\boldsymbol{(R_{4})}$. The construction of $\boldsymbol{(R_{3})}$ guarantees us that it is impossible to create a new caterpillar of type $3$ involving $bufE_{p}(d)$, also $\boldsymbol{(R_{3})}$ is not neutralized. As we assumed a weakly fair daemon, $p$  executes $\boldsymbol{(R_{4})}$ in a finite time. Then, $bufE_{p}(d)$ is empty in a finite time, we are in the case 2.
\end{description}

\end{description}

We can conclude the case 1 by the following affirmation : we are in the case 2 in a finite time.

\item [Case 2:] $bufE_{p}(d)=\varepsilon$.\\
By the definition of a caterpillar of type 1, $p$ is enabled by $\boldsymbol{(R_{2})}$. By the construction of $color_{q}(d)$ and of $\boldsymbol{(R_{2})}$ (for $q$), $\boldsymbol{(R_{2})}$ cannot neutralized for $p$. Since we assumed a weakly fair daemon, we can say that $p$ executes $\boldsymbol{(R_{2}})$ in a finite time. This implies that $C$ disappears and a new caterpillar $C'$ of type 2 associated to $m$ appears. We can now apply the reasoning of the case 1 to deduce that $C'$ becomes a caterpillar of type 1 on $r$ in a finite time.
\end{description}

\end{description}

We have proved that $\mathbf{(P_{\delta})}$ is true, that ended this proof.
\end{proof}

\begin{lemma} \label{lem:depotN}
If routing tables are correct, every processor can generate a first message (\emph{i.e.} it can execute $\boldsymbol{(R_{1})}$) in a finite time under a weakly fair daemon.
\end{lemma}

\begin{proof}
Let $p$ be a processor which has a message $m$ (of Destination $d$) to send. As $p$ has a waiting message, we have $request_{p}=true$ whatever its value in the initial configuration. We must now study two cases:

\begin{description}
\item [Case 1:] $bufR_{p}(d)=\varepsilon$.\\
The processor $p$ executes either $\boldsymbol{(R_{3})}$ or $\boldsymbol{(R_{1})}$ in a finite time (since we assumed a weakly fair daemon and these rules cannot be neutralized).
The result of this execution depends on the value of $choice_{p}(d)$:

\begin{itemize}
\item If $choice_{p}(d)=p$, then $p$ executes $\boldsymbol{(R_{1})}$ in a finite time, we obtain the result.
\item If $choice_{p}(d)\neq p$, then $p$ executes $\boldsymbol{(R_{3})}$ in a finite time. Consequently, $bufR_{p}(d)$ is occupied by a caterpillar of type 3. So, we are in the case 2.1. Note that the fairness of $choice_{p}(d)$ guarantees us that this case cannot appear infinitely. 
\end{itemize}

\item [Case 2:] $bufR_{p}(d)=(m',q,id)$.\\

\begin{description}
\item [Case 2.1:] $bufR_{p}(d)$ belongs to a caterpillar $C$ of type 3.\\
We can apply the reasoning of the case 1.2 of the proof of Lemma \ref{lem:avanceN} to $bufE_{q}(d)$ and conclude that $C$ becomes a caterpillar of type $1$ in a finite time. We are now in the case 2.2.
\item [Case 2.2:] $bufR_{p}(d)$ belongs to a caterpillar $C$ of type 1.\\
We can apply Lemma \ref{lem:avanceN} to $C$ and say that $bufR_{p}(d)$ becomes empty in a finite time. We are now in the case 1.
\end{description}

\end{description}

By the remark of the case 1, this reasoning is finite, that proves the result.
\end{proof}

\begin{lemma} \label{lem:transportN}
If a message $m$ is generated by \AN in a configuration in which routing tables are correct, \AN delivers $m$ to its destination in a finite time under a weakly fair daemon.
\end{lemma}

\begin{proof}
Assume that routing tables are correct when \AN accepts a message $m$ (of Destination $d$) on Processor $p$. This implies that $p$ generated $m$ executing rule $\boldsymbol{(R_{1})}$. This rule leads to the creation of a caterpillar of type 1 associated to $m$ in $bufR_{p}(d)$. Since routing tables are assumed correct and constant, the result follows from $dist(p,d)+1$ applications of Lemma \ref{lem:avanceN}.
\end{proof}

\begin{proposition} \label{prop:snapTRN}
\AN is a snap-stabilizing message forwarding protocol for $\mathcal{SP}'$ if routing tables are correct in the initial configuration.
\end{proposition}

\begin{proof}
Assume that routing tables are correct in the initial configuration. To prove that \AN is a snap-stabilizing message forwarding protocol for specification $\mathcal{SP}'$, we must prove that :

\begin{enumerate}
\item If a processor $p$ requests to send a message, then the protocol is initiated by at least one starting action on $p$ in a finite time. In our case, the starting action is the execution of $\boldsymbol{(R_{1})}$. Lemma \ref{lem:depotN} proves this property. 
\item After a starting action, the protocol is executed according to $\mathcal{SP}'$. If we consider that $\boldsymbol{(R_{1})}$ have been executed at least one time, we can prove that:

\begin{itemize}
\item The first property of $\mathcal{SP}'$ is always satisfied (following Lemma \ref{lem:depotN} and the fact that the waiting for the sending of new messages is blocking). 
\item The second property of $\mathcal{SP}'$ is always satisfied (following Lemma \ref{lem:transportN}). 
\end{itemize}

\end{enumerate}

Consequently, we deduce the proposition.
\end{proof}

\begin{proposition} \label{prop:selfN}
\AN is a self-stabilizing message forwarding protocol for $\mathcal{SP}'$ (even if routing tables are corrupted in the initial configuration) when $\mathcal{A}$ runs simultaneously.
\end{proposition}

\begin{proof}
Remind that $\mathcal{A}$ is a self-stabilizing silent algorithm for computing routing tables running simultaneously to \AN. Moreover, we assumed that $\mathcal{A}$ has priority over \AN (\emph{i.e.} a processor which have enabled actions for both algorithms always chooses the action of $\mathcal{A}$). This guarantees us that routing tables are correct and constant in a finite time regardless of the initial state. 

By Proposition \ref{prop:snapTRN}, \AN is a snap-stabilizing message forwarding protocol for specification $\mathcal{SP}'$ when it starts from a such configuration. Consequently, we obtain the proposition.
\end{proof}

\begin{lemma} \label{lem:perteN}
Under a weakly fair daemon, \AN does not delete a valid message without deliver it to its destination even if $\mathcal{A}$ runs simultaneously.
\end{lemma}

\begin{proof}
By contradiction, let $m$ be a valid message which is deleted without being delivered to its destination.

By the construction of the rule $\boldsymbol{(R_{2})}$, this cannot be the result of an internal forwarding since the message is sequentially copied in $bufE_{p}(d)$ and erased from $bufR_{p}(d)$.

By the construction of rules $\boldsymbol{(R_{5})}$ and $\boldsymbol{(R_{4})}$, this cannot be the result of the execution of $\boldsymbol{(R_{5})}$ (since we are guaranteed that $m$ is in $bufE_{q}(d)$ and cannot be erased from this buffer simultaneously).

By the construction of rules $\boldsymbol{(R_{4})}$ and $\boldsymbol{(R_{2})}$, $m$ cannot be erased from $bufR_{nextHop_{p}(d)}(d)$ in the step in which it is erased from $bufE_{p}(d)$.

Since we have seen that a simultaneous erasing is impossible, the hypothesis implies that $m$ is erased from a buffer $bufE_{p}(d)$ without being copied in another buffer.

The only rule which erases a message from $bufE_{p}(d)$ and does not deliver $m$ is $\boldsymbol{(R_{4})}$. If a processor $p$ executes this rule, then we have $bufE_{p}(d)=(m,q,id)$ and $bufR_{nextHop_{p}(d)}(d)=(m,p,id)$. Assume that the message contained by $bufR_{nextHop_{p}(d)}(d)$ is not the result of the application of rule $\boldsymbol{(R_{3})}$ on $bufE_{p}(d)$. If this message was in $bufR_{nextHop_{p}(d)}(d)$ before $m$ came in $bufE_{p}(d)$, we obtain a contradiction with
the definition of $color_{p}(d)$. This implies that this message came in $bufR_{nextHop_{p}(d)}(d)$ after $m$ came in $bufE_{p}(d)$. Then, the construction of $\boldsymbol{(R_{3})}$ allows us to say that $bufR_{nextHop_{p}(d)}(d)$ contains a message $(m,q',id)$ with $q'\neq p$ (since we have supposed that the message does not come from $bufE_{p}(d)$). We obtain a contradiction. We can conclude that, when we have $bufE_{p}(d)=(m,q,id)$ and $bufR_{nextHop_{p}(d)}(d)=(m,p,id)$, the message $m$  has been copied at least one time. This result contradicts the existence of $m$.
\end{proof}

\begin{lemma} \label{lem:duplicationN}
Under a weakly fair daemon, \AN never duplicates a valid message even if $\mathcal{A}$ runs simultaneously.
\end{lemma}

\begin{proof}
Since the emission of a message creates one caterpillar of type 1 by the construction of the rule $\boldsymbol{(R_{1})}$, it remains to prove the following property : if a caterpillar of type 1 associated to a message $m$ is present on a processor $p$ and this message is erased from all buffers of $p$, then only one neighbor of $p$ contains a caterpillar of type 1 associated to $m$ or $m$ have been delivered to its destination.

Let $C$ be a caterpillar of type 1 associated to a message $m$ (of Destination $d$) on a processor $p$. Since $\boldsymbol{(R_{5})}$ is not enabled for $p$ (by definition of a caterpillar of type 1), $m$ is erased from $bufR_{p}(d)$ by $\boldsymbol{(R_{2})}$. So, $m$ is still present on $p$ (since it has been copied in $bufE_{p}(d)$). Then, we have two cases to observe:

\begin{description}
\item [Case 1:] $p=d$.\\
The only rule for erasing $m$ which can be enabled is $\boldsymbol{(R_{6})}$. This rule delivers $m$ to its destination.
\item [Case 2:] $p\neq d$.\\
The only rule for erasing $m$ which can be enabled is $\boldsymbol{(R_{4})}$. The construction of this rule implies the announced property.
\end{description}

We can conclude that $m$ is delivered at most once to its destination, that proves the result.
\end{proof}

\begin{theorem} \label{th:snapN}
\AN is a snap-stabilizing message forwarding protocol for $\mathcal{SP}$ (even if routing tables are corrupted in the initial configuration) when $\mathcal{A}$ run simultaneously.
\end{theorem}

\begin{proof}
Proposition \ref{prop:selfN} and Lemma \ref{lem:perteN} allows us to conclude that \AN is a snap-stabilizing message forwarding protocol for specification $\mathcal{SP}'$ even if routing tables are corrupted in the initial configuration on condition that $\mathcal{A}$ runs simultaneously.

Then, using this remark and Lemma \ref{lem:duplicationN}, we obtain the result.
\end{proof}

\subsection{Time complexities}\label{sub:analyseN}

Since our algorithm needs a weakly fair daemon, there is no points to do an analysis in terms of steps. It is why all the following complexities analysis are given in rounds. Let $R_{\mathcal{A}}$ be the stabilization time of $\mathcal{A}$ in terms of rounds.

In order to lighten this paper, we present only key ideas of this section proofs.

\begin{proposition} \label{prop:analysemesN}
For any Processor $d$, \AN delivers $2n$ invalid messages to $d$ in the worst case.
\end{proposition}

\begin{sketchproof}
In the initial configuration, the system has at most $2n$ distinct invalid messages of Destination $d$ (since the connected component of the buffer graph associated to $d$ has $2n$ buffers). In the worst case, all these invalid messages are delivered to their destination, that allows us to reach the announced bound.
\end{sketchproof}

\begin{proposition} \label{prop:complexiteN}
In the worst case, a message $m$ (of Destination $d$) needs $O(max(R_{\mathcal{A}},\Delta^{D}))$ rounds to be delivered to $d$ once it has been generated by its source.
\end{proposition}

\begin{sketchproof}
In a first time, we show by induction the following result: if $\gamma$ is a configuration in which routing tables are correct and $C$ is a caterpillar of type 1 associated to a message $m$ (of Destination $d$) on a processor $p$ such as $dist(p,d)=\delta$, then $m$ is delivered to $d$ or there exists a caterpillar of type 1 associated to $m$ on $nextHop_{p}(d)$ in at most $O(\Delta^{\delta})$ rounds. This result is due to the fairness of $choice_{p}(d)$ which can allow at most $\Delta$ messages to ``pass'' $m$ (see the proof of Lemma \ref{lem:avanceN}). 

Then, consider that $s$ is the source of a message $m$ of Destination $d$. We have $dist(s,d)\leq D$ by definition. We can conclude that $m$ is delivered in at most $\underset{\delta=D}{\overset{0}{\sum}}O(\Delta^{\delta})\in O(\Delta^{D})$ rounds if routing tables are correct when $m$ is emitted. 

Finally, we can deduce the result when $m$ is emitted in a configuration in which routing tables are not correct since the message is delivered in at most $O(\Delta^{D})$ rounds after routing tables computation (which takes at most $O(R_{\mathcal{A}})$ rounds if $m$ is not delivered during the routing tables computation since we have assumed the priority of $\mathcal{A}$ over \AN).
\end{sketchproof}

\begin{proposition} \label{prop:delaiN}
The delay (waiting time before the first emission) and the waiting time (between two consecutive emissions) of \AN is $O(max(R_{\mathcal{A}},\Delta^{D}))$ rounds in the worst case.
\end{proposition}

\begin{sketchproof}
Let $p$ be a processor which has a message of Destination $d$ to emit. By the fairness of $choice_{p}(d)$, we can say that $m$ is generated after at most $(\Delta-1)$ releases of $bufR_{p}(d)$ (see proof of Lemma \ref{lem:avanceN}). The result of Proposition \ref{prop:complexiteN} allows us to say that $bufR_{p}(d)$ is released in $O(max(R_{\mathcal{A}},\Delta^{D}))$ rounds at worst. Indeed, we can deduce the result.
\end{sketchproof}

The complexity obtained in Proposition \ref{prop:complexiteN} is due to the fact that the system delivers a huge quantity of messages during the forwarding of the considered message. It's why we interest now in the amortized complexity (in rounds) of our algorithm. For an execution $\Gamma$, this measure is equal to the number of rounds of $\Gamma$ divided by the number of delivered messages during $\Gamma$ (see \cite{CLRS02} for a formal definition).

\begin{proposition} \label{prop:amortieN}
The amortized complexity (to forward a message) of \AN is $O(max$ $(R_{\mathcal{A}},D))$ rounds.
\end{proposition}

\begin{sketchproof}
In a first time, we must prove the following property: if $\gamma$ is a configuration in which at least one message of Destination $d$ is present and in which routing tables are correct, then \AN delivers at least one message to $d$ in the $3D$ rounds following $\gamma$. 

The proof of this property is done as follows. Let $\delta$ be the smallest number such that there exists a message of Destination $d$ on a processor $p$ which satisfy $dist(p,d)=\delta$. Then, we prove that, after at most three rounds, there exists a message (not necessarily $m$) on a processor $p'$ which satisfies $dist(p',d)=\delta-1$. Since $\delta\leq D$ in $\gamma$, we obtain the announced property.
 
Assume now an initial configuration in which routing tables are correct. Let $\Gamma$ be one execution leads to the worst amortized complexity. Let $R_{\Gamma}$ be the number of rounds of $\Gamma$. By the previous property, we can say that \AN delivers at least $\frac{R_{\Gamma}}{3D}$ messages during $\Gamma$. So, we have an amortized complexity of $\frac{R_{\Gamma}}{\frac{R_{\Gamma}}{3D}}\in\Theta(D)$. Then, the announced result is obvious.
\end{sketchproof}

\subsection{Conclusion}

In this section, we prove that we can adapt the ``destination-based'' deadlock-free controller defined in \cite{MS78} to obtain a snap-stabilizing message forwarding algorithm. Our algorithm is mainly based on the control of effects of routing tables moves on message. This control is performed in two ways. Firstly, we ``slow down'' messages by using two buffers per processor in order to control the number of copy of a same message in the network at a given time. Secondly, we use a specific flag to avoid message merge or duplication.

The initial fault-free protocol uses $n^{2}$ buffers for the whole network and our protocol uses $2n^{2}$ buffers. Consequently, our protocol ensures a stronger safety and fault-tolerance with respect the initial one without a significant overcost in space. Our time analysis (see Section \ref{sub:analyseN}) shows that this stronger safety does not leads to an overcost in time.
 
\section{Second protocol}\label{sec:protocolD}

\subsection{Informal description}

In this section, we give a second snap-stabilizing message forwarding protocol adapted to the ``distance-based'' deadlock-free controller (see Section \ref{sec:survey}). Our idea is to adapt this scheme in order to tolerate transient faults. To perform this goal, we assume the existence of a self-stabilizing silent (\emph{i.e.} no actions are enabled after convergence) algorithm $\mathcal{A}$ to compute routing tables which runs simultaneously to our message forwarding protocol. Moreover, we assume that $\mathcal{A}$ has priority over our protocol (\emph{i.e.} a processor which has enabled actions for both algorithms always chooses the action of $\mathcal{A}$). This guarantees us that routing tables are correct and constant in a finite time. To simplify the presentation, we assume that $\mathcal{A}$ induces only minimal paths in number of edges. We assume that our protocol can have access to the routing table via a function, called $nextHop_{p}(d)$. This function returns the identity of the neighbor of $p$ to which $p$ must forward messages of Destination $d$. 

Our idea is as follows. We choose exactly the same graph buffer as \cite{MS78} and we allow the erasing of a message only if we are assured that the message has been delivered to its destination. In this goal, we use an acknowledgment scheme which guarantees the reception of the message.

More precisely, we associate to each copy of the message a type which has 3 values: $S$ (Sending), $A$ (Acknowledgment) and $F$ (Fail). Forwarding of a valid message follows the above scheme:

\begin{enumerate}
\item Generation with type $S$ in a buffer of rank $1$.
\item Forwarding (with copy in buffers of increasing rank) with type $S$ without any erasing.
\item If the message reaches its destination :
\begin{enumerate}
\item It is delivered and the copy of the message takes type $A$.
\item Type $A$ is propagated to the sink of the message following the income path.
\item Buffers are allowed to free themselves once the type $A$ is propagated to the previous buffer on the path.
\item The sink erases its copy, that performs the erasing of the message.
\end{enumerate}
\item Otherwise, (the message reaches a buffer of rank $D+1$ without cross its destination) :
\begin{enumerate}
\item The copy of the message takes type $F$.
\item Type $F$ is propagated to the sink of the message following the income path.
\item Buffers are allowed to free themselves once the type $F$ is propagated to the previous buffer on the path.
\item Then, the sink of the message gives the type $S$ to its copy, that begin a new cycle (the message is sending once again).
\end{enumerate}
\end{enumerate}

Obviously, it is necessary to take in account invalid messages: we have chosen to let them follow the forwarding scheme and to erase them if they reach step 4.d.

The key idea of the snap-stabilization of our algorithm is the following: since a valid message is never erased, it is sent again after the stabilization of routing tables (if it never reached its destination before) and it is then normally forwarded. 

To avoid livelocks, we use a fair scheme of selection of processors allowed to forward a message for each buffer. We can manage this fairness by a queue of requesting processors. Finally, we use a specific flag to prevent message losses. It is composed of the identity of the next processor on the path of the message, the identity of the last processor cross over by the message, the identity of the destination of the message and the type of the message ($S$, $A$ or $F$).

We must manage a communication between our algorithm and processors in order to know when a processor has a message to send. We have chosen to create a Boolean shared variable $request_{p}$ (for any processor $p$). Processor $p$ can set it at $true$ when it is at $false$ and when $p$ has a message to send. Otherwise, $p$ must wait that our algorithm sets the shared variable to $false$ (when a message is sent out).

\subsection{Algorithm}

We now present formally our protocol in Algorithm \ref{algo:AD}. We call it \AD for $\mathcal{S}$nap-$\mathcal{S}$tabilizing $\mathcal{M}$essage $\mathcal{F}$orwarding $\mathcal{P}$rotocol 2.

\begin{algorithm}
	\caption{\label{algo:AD}\AD : Message forwarding protocol for processor $p$.}
\small{
		\textbf{Data:}\\
			- $n,D$ : natural numbers equal resp. to the number of processors and to the diameter of the network.\\
			- $I=\{0,...,n-1\}$ : set of processor identities of the network.\\
			- $N_{p}$ : set of neighbors of $p$.\\
		\textbf{Message:}\\
			- $(m,r,q,d,c)$ with $m$ useful information of the message, $r\in N_{p}$
			identity of the next processor to cross for the message (when it reaches the node), 
			$q\in N_{p}$ identity of the last processor cross over by the message, $d\in I$ 
			identity of the destination of the message, $c\in\{S,A,F\}$ type of the message.\\
		\textbf{Variables:}\\
			- $\forall i\in\{1,...,D+1\}$, $buf_{p}(i)$ : buffer which can contain a message or be empty (denoted by $\varepsilon$)\\
		\textbf{Input/Output:}\\
			- $request_{p}$ : Boolean. The higher layer can set it to "true" when its value is "false"
			and when there is a waiting message. We consider that this waiting is blocking.\\
			- $nextMes_{p}$: gives the message waiting in the higher layer.\\
			- $nextDest_{p}$: gives the destination of $nextMes_{p}$ if it exists, $null$ otherwise.\\
		\textbf{Procedures:}\\
			- $nextHop_{p}(d)$: neighbor of $p$ given by the routing for Destination $d$ (if $d=p$, we choose arbitrarily $r\in N_{p}$).\\
			- $\forall i\in \{2,...,D+1\},choice_{p}(i)$: fairly chooses one of the processors which can send a message
			in $buf_{p}(i)$, \emph{i.e.} $choice_{p}(d)$ satisfies predicate 
			$((choice_{p}(i)\in N_{p}) \wedge (buf_{choice_{p}(i)}(i-1)=(m,p,q,d,S)) \wedge (choice_{p}(i)\neq d))$. 
			We can manage this fairness with a queue of length $\Delta+1$ of processors which satisfies the
			predicate.\\
			- $deliver_{p}(m)$: delivers the message $m$ to the higher layer
			of $p$.\\
		\textbf{Rules:}
			\begin{description}
				\item \textbf{/*} Rules for the buffer of rank $1$ \textbf{*/}\\
					\textbf{/*} Generation of messages \textbf{*/}\\
					$\boldsymbol{(R_{1})}$ :: $request_{p} \wedge (buf_{p}(1)=\varepsilon) \wedge 
					(nextDest_{p}=d) \wedge (nextMes_{p}=m) \wedge (buf_{nextHop_{p}(d)}(2)\neq(m,r',p,d,c))
					\longrightarrow buf_{p}(1):=(m,nextHop_{p}(d),r,d,S)$ with $r\in N_{p};\;	request_{p}:=false$\\
					\textbf{/*} Processing of acknowledgment \textbf{*/}\\
					$\boldsymbol{(R_{2})}$ :: $(buf_{p}(1)=(m,r,q,d,F)) \wedge (d\neq p)
					\wedge (buf_{r}(2)\neq(m,r',p,d,F)) \longrightarrow buf_{p}(1):=(m,nextHop_{p}(d),q,d,S)$ \\
					$\boldsymbol{(R_{3})}$ :: $(buf_{p}(1)=(m,r,q,d,A)) \wedge (d\neq p) \wedge (buf_{r}(2)\neq(m,r',p,d,A)) \longrightarrow buf_{p}(1):=\varepsilon$ \\
					\textbf{/*} Management of messages which reach their destinations \textbf{*/}\\
					$\boldsymbol{(R_{4})}$ :: $buf_{p}(1)=(m,r,q,p,S)\longrightarrow deliver_{p}(m);\; buf_{p}(1):=(m,r,q,p,A)$ \\
					$\boldsymbol{(R_{5})}$ :: $buf_{p}(1)=(m,r,q,p,A)\longrightarrow buf_{p}(1):=\varepsilon$ \\
					$\boldsymbol{(R_{6})}$ :: $buf_{p}(1)=(m,r,q,p,F)\longrightarrow buf_{p}(1):=(m,r,q,p,S)$
				\item \textbf{/*} Rule for buffers of rank $1$ to $D$ : propagation of acknowledgment \textbf{*/}\\
					$\boldsymbol{(R_{7})}$ :: $\exists i\in\{1,...,D\},((buf_{p}(i)=(m,r,q,d,S)) \wedge (p\neq d) \wedge (buf_{r}(i+1)=(m,r',p,d,c)) \wedge$
					$(c\in\{R,A\})) \longrightarrow buf_{p}(i):=(m,r,q,d,c)$ 
				\item \textbf{/*} Rules for buffers of rank $2$ to $D$ \textbf{*/}\\
					\textbf{/*} Forwarding of messages \textbf{*/}\\
					$\boldsymbol{(R_{8})}$ :: $\exists i\in\{2,...,D\},( (buf_{p}(i)=\varepsilon) \wedge (choice_{p}(i)=s) \wedge (buf_{s}(i-1)=(m,p,q,d,S)) \wedge$
					$(buf_{nextHop_{p}(d)}(i+1)\neq(m,r,p,d,c))) \longrightarrow buf_{p}(i):=(m,nextHop_{p}(d),s,d,S)$\\
			\textbf{/*} Erasing of messages of which the acknowledgment has been forwarded \textbf{*/}\\
				$\boldsymbol{(R_{9})}$ :: $\exists i\in\{2,...,D\},((buf_{p}(i)=(m,r,q,d,c)) \wedge (c\in\{F,A\}) \wedge (d\neq p)
				\wedge (buf_{q}(i-1)=(m,p,q',d,c)) \wedge (buf_{r}(i+1)\neq(m,r',p,d,c)) \longrightarrow buf_{p}(i):=\varepsilon$
		\item \textbf{/*} Rules for buffers of rank $2$ to $D+1$ \textbf{*/}\\
			\textbf{/*} Consumption of a message and generation of the acknowledgment $A$ \textbf{*/}\\
			$\boldsymbol{(R_{10})}$ :: $\exists i\in\{2,...,D+1\},buf_{p}(i)=(m,r,q,p,S)\longrightarrow deliver_{p}(m);\;buf_{p}(i):=(m,r,q,p,A)$ \\
	\textbf{/*} Erasing of messages of destination $p$ of which the acknowledgment has been forwarded \textbf{*/}\\
			$\boldsymbol{(R_{11})}$ :: $\exists i\in\{2,...,D+1\},((buf_{p}(i)=(m,r,q,p,c)) \wedge (c\in\{F,A\}) \wedge (buf_{q}(i-1)=(m,p,q',p,c)))\longrightarrow buf_{p}(i):=\varepsilon$
			\end{description}
}
\end{algorithm}

\begin{algorithm}
	\begin{description}
		\item \textbf{End of Algorithm \ref{algo:AD}:}
\small{
			\item \textbf{/*} Rules for the buffer of rank $D+1$ \textbf{*/}\\
			\textbf{/*} Forwarding of messages \textbf{*/}\\
			$\boldsymbol{(R_{12})}$ :: $(buf_{p}(D+1)=\varepsilon) \wedge (choice_{p}(D+1)=s) \wedge (buf_{s}(D)=(m,p,q,d,S))\longrightarrow buf_{p}(D+1):=(m,nextHop_{p}(d),s,d,S)$ 
			\textbf{/*} Generation of the acknowledgment $F$ \textbf{*/}\\
			$\boldsymbol{(R_{13})}$ :: $(buf_{p}(D+1)=(m,r,q,d,S)) \wedge (d\neq p)\longrightarrow buf_{p}(D+1):=(m,r,q,d,F)$ \\
			\textbf{/*} Erasing of messages of which the acknowledgment has been forwarded \textbf{*/}\\
			$\boldsymbol{(R_{14})}$ :: $(buf_{p}(D+1)=(m,r,q,d,c)) \wedge (c\in\{F,A\}) \wedge (d\neq p) \wedge (buf_{q}(D)=(m,p,q',d,c))\longrightarrow buf_{p}(D+1):=\varepsilon$ 
		\item \textbf{/*} Correction rules: erasing of tail of abnormal caterpillars of type $F,A$ (\emph{cf.} definitions below) \textbf{*/}\\
			$\boldsymbol{(R_{15})}$ :: $\exists i\in\{2,...,D\},((buf_{p}(i)=(m,r,q,d,c)) \wedge (c\in\{F,A\})
			\wedge (buf_{r}(i+1)\neq(m,r',p,d,c)) \wedge (buf_{q}(i-1)\neq(m,p,q',d,c')))\longrightarrow buf_{p}(i):=\varepsilon$ \\
			$\boldsymbol{(R_{16})}$ :: $\exists i\in\{2,...,D\},((buf_{p}(i)=(m,r,q,d,c)) \wedge (c\in\{F,A\}) \wedge (buf_{r}(i+1)\neq(m,r',p,d,c)) 
			\wedge (buf_{q}(i-1)=(m,p,q',d,c')) \wedge (c'\in\{F,A\}\backslash\{c\} \vee q=d)) \longrightarrow buf_{p}(i):=\varepsilon$ \\
			$\boldsymbol{(R_{17})}$ :: $(buf_{p}(D+1)=(m,r,q,d,c)) \wedge (c\in\{F,A\}) \wedge (buf_{q}(D)\neq(m,p,q',d,c'))\longrightarrow buf_{p}(D+1):=\varepsilon$ \\
			$\boldsymbol{(R_{18})}$ :: $(buf_{p}(D+1)=(m,r,q,d,c)) \wedge (c\in\{F,A\}) \wedge (buf_{q}(D)=(m,p,q',d,c')) \wedge (c'\in\{F,A\}\backslash\{c\} \vee q=d)\longrightarrow buf_{p}(D+1):=\varepsilon$
}
	\end{description}
\end{algorithm}

\subsection{Proof of correctness}

In order to simplify the proof, we introduce a second specification of the problem. This specification allows message duplications.
      
\begin{specification} [$\mathcal{SP}'$]
Specification of message forwarding problem allowing duplication.
\begin{itemize}
\item Any message can be send out in a finite time.
\item Any valid message is delivered to its destination in a finite time.
\end{itemize}
\end{specification}

In this section, we prove that \AD is a snap-stabilizing message forwarding protocol for specification $\mathcal{SP}$. For that, we are going to prove successively that:

\begin{enumerate}
\item Copies of a same message have a particular structure. Then, we prove some properties on the behavior of these structures under \AD(Lemmas \ref{lem:prelem1}, \ref{lem:prelem2}, \ref{lem:prelem3}, and \ref{lem:prelem4}).
\item \AD is a snap-stabilizing message forwarding protocol for specification $\mathcal{SP}'$ if routing tables are correct in the initial configuration (Lemmas \ref{lem:chenilleSA}, \ref{lem:depotD}, \ref{lem:transportD} and Proposition \ref{prop:snapTRD}). 
\item \AD is a self-stabilizing message forwarding protocol for specification $\mathcal{SP}'$ even if routing tables are corrupted in the initial configuration (Proposition \ref{prop:selfD}).
\item \AD is a snap-stabilizing message forwarding protocol for specification $\mathcal{SP}$ even if routing tables are corrupted in the initial configuration (Lemmas \ref{lem:perteD}, \ref{lem:duplicationD} and Theorem \ref{th:snapD}).
\end{enumerate}

In this proof, we consider that the notion of message is different from the notion of useful information. This implies that two messages with the same useful information sent by the same processor are considered as two different messages. We must prove that the algorithm does not loose one of them thanks to the use of the flag.

\paragraph{Preliminaries.}In a first time, we define a particular structure of messages and we study the behavior of these structure under \AD. Let $\gamma$ be a configuration of the network. We say that a message $m$ is existing in $\gamma$ if at least one buffer contains $m$ in $\gamma$.

\begin{definition} [Caterpillar of a message $m$] 
Let $m$ be a message of Destination $d$ existing in a configuration $\gamma$. We define a caterpillar associated to $m$ (noted $C_{m}$) as the longest sequence of buffers $C_{m}=buf_{p_{1}}(i)...buf_{p_{t}}(i+t-1)$ (with $t \geq 1$) which satisfies:
\begin{itemize}
\item $\forall j\in\{1,...,t-1\}$, $p_{j}\neq d$ and $p_{j+1}\neq p_{j}$.
\item $\forall j\in\{1,...,t\}$, $buf_{p_{j}}(i+j-1)=(m,r_{j},q_{j},d,c_{j})$.
\item $\forall j\in\{1,...,t-1\}$, $r_{j}=p_{j+1}$.
\item $\forall j\in\{2,...,t\}$, $q_{j}=p_{j-1}$.
\item $\exists k\in\{1,...,t+1\}$, $\begin{cases}
\forall j\in\{1,...,k-1\},\; c_{j}=S\; and\\
(\forall j\in\{k,...,t\},\; c_{j}=A)\vee(\forall j\in\{k,...,t\},\; c_{j}=F)\end{cases}$
\end{itemize}
We call respectively $buf_{p_{1}}(i)$, $buf_{p_{t}}(i+t-1)$, and $lg_{C_{m}}=t$ the tail, the head, and the length of $C_{m}$.
\end{definition}

We give now some characterization for caterpillars.

\begin{definition} [Characterization of caterpillar of a message $m$]
Let $m$ be a message of Destination $d$ in a configuration $\gamma$ and $C_{m}=buf_{p_{1}}(i)...buf_{p_{t}}(i+t-1)$ ($t\geq1$) a caterpillar associated to $m$. Then,
\begin{itemize}
\item $C_{m}$ is a normal caterpillar if $i=1$. It is abnormal otherwise ($i\geq2$).
\item $C_{m}$ is a caterpillar of type $S$ if $\forall j\in\{1,...,t\}$, $c_{j}=S$ (\emph{i.e.} $k=t+1$).
\item $C_{m}$ is a caterpillar of type $A$ if $\exists j\in\{1,...,t\}$, $c_{j}=A$ (\emph{i.e.} $k<t+1$).
\item $C_{m}$ is a caterpillar of type $F$ if $\exists j\in\{1,...,t\}$, $c_{j}=F$ (\emph{i.e.} $k<t+1$).
\end{itemize}
\end{definition}

\begin{figure}
\begin{centering}
\includegraphics[scale=0.4]{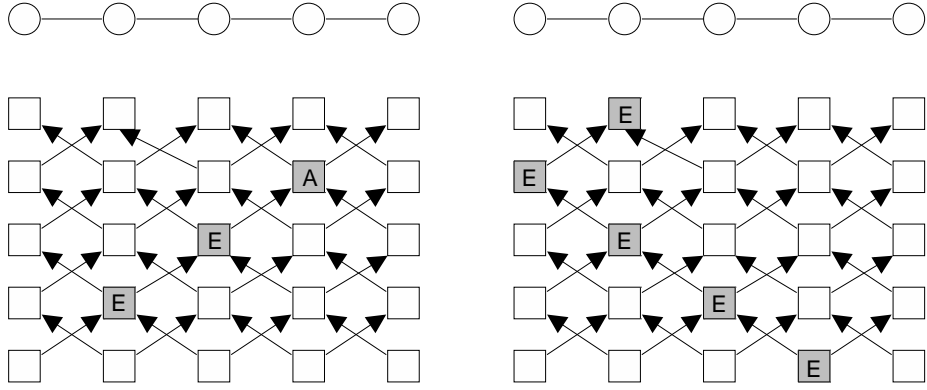}
\par\end{centering}
\caption{\label{fig:CaterpillarD}Examples of caterpillar (at left: abnormal of type $A$, at right: normal of type $E$).}
\end{figure}

It is obvious that, for each caterpillar $C_{m}$, either $C_{m}$ is normal or abnormal. In the same way, $C_{m}$ is only of type $S$, $A$ or $F$. The reader can find in Figure \ref{fig:CaterpillarD} an example for some type of caterpillar. 

\begin{lemma} \label{lem:prelem1}
Let $\gamma$ be a configuration and $m$ be a message of Destination $d$ existing in $\gamma$. Under a weakly fair daemon, every abnormal caterpillar of type $F$ (resp. $A$) associated to $m$ disappears in a finite time or become a normal caterpillar of type $F$ (resp. $A$).
\end{lemma}

\begin{proof}
Let $\gamma$ be a configuration of the network. Let $m$ be an existing message (of Destination $d$) in $\gamma$. Let $C_{m}=buf_{p_{1}}(i)...buf_{p_{t}}(i+t-1)$ ($t\geq1$ and $i>1$) be a normal caterpillar of type $F$ or $A$  associated to $m$. Let $c$ be the type of $C_{m}$.

\begin{enumerate}
\item By definition of caterpillar of type $c$, we have $1\leq k\leq t$. We can deduce that $i+k-2<i+t-1\leq D+1$ and then $\boldsymbol{(R_{7})}$ is enabled for $p_{k-1}$. This rule can not be neutralized since Processor $p_{k}$ is not enabled by a rule affecting its buffer of rank $i+k$. As the daemon is weakly fair, $p_{k-1}$ executes these rule in a finite time. We can repeat this reasoning $k-1$ times on Processors $p_{k-1},...,p_{1}$. Then, we obtain a caterpillar which all buffers are on type $c$ in a finite time.
\item If $t=1$, we can directly go to case 4. Otherwise ($t\geq2$), we must distinguish the following cases:

\begin{description}
\item [Case 1:] $p_{t}=d$.\\
Processor $p_{t}$ is the enabled for rule $\boldsymbol{(R_{11})}$ by definition of a caterpillar and the fact that all buffers of $C_{m}$ are of type $c$. Note that Processor $p_{t-1}$ is not enabled. Consequently, this rule remains infinitely enabled for $p_{t}$. Since the daemon is weakly fair, $p_{t}$ executes this rule in a finite time. Then, $buf_{p_{t}}(i+t-1)$ is empty in a finite time.
\item [Case 2:] $p_{t}\neq d$.\\

\begin{description}
\item [Case 2.1:] $i+t-1=D+1$.\\
Then, Processor $p_{t}$ is enabled for rule $\boldsymbol{(R_{14})}$ by definition of a caterpillar and the fact that all buffers of $C_{m}$ are of type $c$. Note that Processor $p_{t-1}$ is not enabled. Consequently, this rule remains infinitely enabled for $p_{t}$. Since the daemon is weakly fair, $p_{t}$ executes this rule in a finite time. Then, $buf_{p_{t}}(i+t-1)$ is empty in a finite time.
\item [Case 2.2:] $i+t-1\leq D$.\\
Assume that $buf_{p_{t}}(i+t-1)=(m,r,q,d,c)$. Then, Processor $p_{t}$ is enabled for rule $\boldsymbol{(R_{9})}$ by definition of a caterpillar and the fact that all buffers of $C_{m}$ are of type $c$. Note that Processor $p_{t-1}$ is not enabled and that Processor $r$ cannot forward a message $(m,r',p_{t},d,c)$ in its buffer of rank $i+t$ (since $buf_{p_{t}}(i+t-1)$ is of type $c\neq S$). Consequently, this rule remains infinitely enabled for $p_{t}$. Since the daemon is weakly fair, $p_{t}$ executes this rule in a finite time. Then, $buf_{p_{t}}(i+t-1)$ is empty in a finite time.
\end{description}

\end{description}

\item By following a reasoning similar to the one of case 2.2, we can prove that $p_{t-1},...,p_{2}$ executes $\boldsymbol{(R_{9})}$ sequentially in a finite time
\item Then, we obtain a caterpillar of type $c$ of length $1$ satisfying $i>1$. Assume that $buf_{p_{1}}(i)=(m,r,q,d,c)$. We can distinguish the following cases:

\begin{description}
\item [Case 1:] $buf_{q}(i-1)=(m,p_{1},q',d,c')$.

\begin{description}
\item [Case 1.1:]  $q=d$.\\
By the definition of a caterpillar of type $c$ of length $1$ and the hypothesis, $p_{1}$ is enabled for rule $\boldsymbol{(R_{16})}$ (if $i\leq D$) or $\boldsymbol{(R_{18})}$ (if $i=D+1$). By a reasoning similar to the one of case 2.2 above, these rule remains infinitely enabled. Since the daemon is weakly fair, $p_{1}$ executes this rule in a finite time. Consequently, $buf_{p_{1}}(i)$ becomes empty in a finite time. Then, $C_{m}$ disappears.
\item [Case 1.2:] $q\neq d$.\\
Assume that $c'=S$. Then, $buf_{q}(i-1)$ belongs to $C_{m}$. This contradicts the fact that $C_{m}$ is of type $c$. Consequently, $c'\in\{F,A\}$. 

If $c'=c$, then the execution of rule $\boldsymbol{(R_{7})}$ by $p_{1}$ leads to the merge of two caterpillars of type $c$. Then, consider the new caterpillar  $C'_{m}=buf_{p'_{1}}(i')...buf_{p'_{t'}}(i'+t'-1)$ (with $buf_{p'_{t'}}(i'+t'-1)=buf_{p_{1}}(i)$). If $i'=1$, then we have a normal caterpillar of type $c$. Otherwise, we can restart the reasoning (we are ensured that this reasoning is finite since we have $1\leq i'<i$ at each step).

Consider now the case $c'\neq c$. By definition of a caterpillar of type $c$ of length $1$ and the hypothesis, $p_{1}$ is enabled by rule $\boldsymbol{(R_{16})}$ (if $i\leq D$) or $\boldsymbol{(R_{18})}$ (if $i=D+1$). By a reasoning similar to the one of case 2.2 above, these rule remains infinitely enabled. Since the daemon is weakly fair, $p_{1}$ executes this rule in a finite time. Consequently, $buf_{p_{1}}(i)$ becomes empty in a finite time. Then, $C_{m}$ disappears.
\end{description}

\item [Case 2:] $buf_{q}(i-1)\neq(m,p_{1},q',d,c')$.\\
By definition of a caterpillar of type $c$ of length $1$ and the hypothesis, $p_{1}$ is enabled by rule $\boldsymbol{(R_{15})}$ (if $i\leq D$) or $\boldsymbol{(R_{17})}$ (if $i=D+1$). By a reasoning similar to the one of case 2.2 above, these rule remains infinitely enabled. Since the daemon is weakly fair, $p_{1}$ executes this rule in a finite time. Consequently, $buf_{p_{1}}(i)$ becomes empty in a finite time. Then, $C_{m}$ disappears.
\end{description}

\end{enumerate}

In all cases, $C_{m}$ disappears or becomes a normal caterpillar of type $c$ in a finite time, that leads us to the lemma.
\end{proof}

\begin{lemma} \label{lem:prelem2}
Let $\gamma$ be a configuration and $m$ be a message of Destination $d$ existing in $\gamma$. Under a weakly fair daemon, every normal caterpillar of type $A$ associated to $m$ disappears in a finite time.
\end{lemma}

\begin{proof}
Let $\gamma$ be a configuration and $m$ be a message of Destination $d$ existing in $\gamma$. Let $C_{m}=buf_{p_{1}}(1)...buf_{p_{t}}(t)$ ($t\geq1$) be a normal caterpillar of type $A$ associated to $m$. We must distinguish the following cases:

\begin{description}
\item [Case 1:] $t=1$.

\begin{description}
\item [Case 1.1:] $p_{1}=d$.\\
Then, rule $\boldsymbol{(R_{5})}$ is enabled for $p_{1}$. Since the guard of this rule involves only local variables, it remains infinitely enabled. Since the daemon is weakly fair, $p_{1}$ executes this rule in a finite time. Consequently, $C_{m}$ disappears.
\item [Case 1.2:] $p_{1}\neq d$.\\
By the definition of a caterpillar and the hypothesis, $p_{1}$ is enabled by rule $\boldsymbol{(R_{3})}$. By a reasoning similar to the one of the case 2.2.2 of the proof of Lemma \ref{lem:prelem1}, we can prove that this rule remains infinitely enabled. Since the daemon is weakly fair, $p_{1}$ executes this rule in a finite time. Consequently, $C_{m}$ disappears.
\end{description}

\item [Case 2:] $t\geq2$.\\
We can apply the reasoning of points 1,2, and 3 of the proof of Lemma \ref{lem:prelem1}. That leads us to case 1.2. 
\end{description}

In all the cases, $C_{m}$ disappears in a finite time, that leads us to the lemma.
\end{proof}

\begin{lemma} \label{lem:prelem3}
Let $\gamma$ be a configuration and $m$ be a message of Destination $d$ existing in $\gamma$. Under a weakly fair daemon, every normal caterpillar of type $F$ associated to $m$ becomes a normal caterpillar of type $S$ of length $1$ in a finite time.
\end{lemma}

\begin{proof}
Let $\gamma$ be a configuration and $m$ be a message of Destination $d$ existing in $\gamma$. Let $C_{m}=buf_{p_{1}}(1)...buf_{p_{t}}(t)$ ($t\geq1$) be a normal caterpillar of type $F$ associated to $m$. We must distinguish the following cases:

\begin{description}
\item [Case 1:] $t=1$.

\begin{description}
\item [Case 1.1:] $p_{1}=d$.\\
Then, rule $\boldsymbol{(R_{6})}$ is enabled for $p_{1}$. Since the guard of this rule involves only local variables, it remains infinitely enabled. Since the daemon is weakly fair, $p_{1}$ executes this rule in a finite time. Consequently, $C_{m}$ becomes a caterpillar of type $S$ of length $1$.
\item [Case 1.2:] $p_{1}\neq d$.\\
By the definition of a caterpillar and the hypothesis, $p_{1}$ is enabled by rule $\boldsymbol{(R_{2})}$. By a reasoning similar to the one of the case 2.2.2 of the proof of Lemma \ref{lem:prelem1}, we can prove that this rule remains infinitely enabled. Since the daemon is weakly fair, $p_{1}$ executes this rule in a finite time. Consequently, $C_{m}$ becomes a caterpillar of type $S$ of length $1$.
\end{description}

\item [Case 2:] $t\geq2$.\\
We can apply the reasoning of points 1,2, and 3 of the proof of Lemma \ref{lem:prelem1}. That leads us to case 1.2. 
\end{description}

In all cases, we proved that $C_{m}$ becomes a caterpillar of type $S$ of length $1$ in a finite time, that leads us to the lemma.
\end{proof}

\begin{lemma} \label{lem:prelem4}
Let $\gamma$ be a configuration and $m$ be a message of Destination $d$ existing in $\gamma$. Under a weakly fair daemon, every caterpillar of type $S$ associated to $m$ becomes a caterpillar of type $A$ or $F$ in a finite time.
\end{lemma}

\begin{proof}
Let $\gamma$ be a configuration of the network and $m$ be a message (of Destination $d$) existing in $\gamma$. Let $C_{m}=buf_{p_{1}}(i)...buf_{p_{t}}(i+t-1)$ ($t\geq1$) be a caterpillar of type $S$ associated to $m$.

We prove this result by a decreasing induction on the rank of the buffer occupied by the head of $C_{m}$ in $\gamma$. Let us define the following property:

$\boldsymbol{(P_{l})}$ : If $C_{m}$ satisfies $i+t-1=l$, then it becomes a caterpillar of type $A$ or $F$ in a finite time.

\begin{description}
\item [Initialization:] We want to prove that $\boldsymbol{(P_{D+1})}$ is true.\\
Let $C_{m}=buf_{p_{1}}(i)...buf_{p_{t}}(i+t-1)$ ($t\geq1$) be a caterpillar of type $S$ associated to $m$ such that $i+t-1=D+1$. We must distinguish the following cases:

\begin{description}
\item [Case 1:] $p_{t}=d$.\\
By hypothesis, Processor $p_{t}$ is enabled for rule $\boldsymbol{(R_{10})}$. Since the guard of this rule involves only local variables, it remains infinitely enabled. Since the daemon is weakly fair, $p_{t}$ executes this rule in a finite time. Consequently, $buf_{p_{t}}(i+t-1)$ becomes a buffer of type $A$ and $C_{m}$ becomes a caterpillar of type $A$ in a finite time. Then, Property $\boldsymbol{(P_{D+1})}$ is satisfied.
\item [Case 2:] $p_{t}\neq d$.\\
By hypothesis, Processor $p_{t}$ is enabled for rule $\boldsymbol{(R_{13})}$. Since the guard of this rule involves only local variables, it remains infinitely enabled. Since the daemon is weakly fair, $p_{t}$ executes this rule in a finite time. Consequently, $buf_{p_{t}}(i+t-1)$ becomes a buffer of type $F$ and $C_{m}$ becomes a caterpillar of type $F$ in a finite time. Then, Property $\boldsymbol{(P_{D+1})}$ is satisfied.
\end{description}

\item [Induction:] Let be $l\leq D$. Assume that $\boldsymbol{(P_{l+1})}...\boldsymbol{(P_{D+1})}$ are satisfied. We want to prove that $\boldsymbol{(P_{l})}$ is then satisfied.\\
Let $C_{m}=buf_{p_{1}}(i)...buf_{p_{t}}(i+t-1)$ ($t\geq1$) be a caterpillar of type $S$ associated to $m$ such that $i+t-1=l<D+1$. We must distinguish the following cases:

\begin{description}
\item [Case 1:] $p_{t}=d$.

\begin{description}
\item [Case 1.1:] $i+t-1=1$.\\
By hypothesis, Processor $p_{t}$ is enabled for rule $\boldsymbol{(R_{4})}$. Since the guard of this rule involves only local variables, it remains infinitely enabled. Since the daemon is weakly fair, $p_{t}$ executes this rule in a finite time. Consequently, $buf_{p_{t}}(i+t-1)$ becomes a buffer of type $A$ and $C_{m}$ becomes a caterpillar of type $A$ in a finite time. Then, Property $\boldsymbol{(P_{l})}$ is satisfied.
\item [Case 1.2:] $2\leq i+t-1\leq D$.\\
These case is similar to the case 1 of initialization. Consequently, $C_{m}$ becomes a caterpillar of type $A$ in a finite time. Then, Property $\boldsymbol{(P_{l})}$ is satisfied.
\end{description}

\item [Case 2:] $p_{t}\neq d$.\\

Assume w.l.g. that $buf_{p_{t}}(i+t-1)=(m,r,q,d,E)$. We want to prove that the head of $C_{m}$ goes up of one buffer in a finite time. We must study the following cases:

\begin{description}
\item [Case 2.1:] $i+t=D+1$.\\

\begin{enumerate}
\item If $buf_{r}(i+t)=\varepsilon$, then Processor $r$ is enabled by rule $\boldsymbol{(R_{12})}$. Since Processor $choice_{r}(i+t)$ is not enabled, this rule remains infinitely enabled for $r$. Processor $r$ executes this rule in a finite time because the daemon is weakly fair. The result of this execution depends on the value of $choice_{r}(i+t)$:

\begin{enumerate}
\item If $choice_{r}(i+t)=p_{t}$, then the head of $C_{m}$ goes up of one buffer when $r$ executes rule $\boldsymbol{(R_{12})}$.
\item If $choice_{r}(i+t)=s\neq p_{t}$, then $buf_{r}(i+t)$ takes the value $(m',r',s,d',c)$ when $r$ executes rule $\boldsymbol{(R_{12})}$. This leads us to case 2.b. Note that the fairness of  $choice_{r}(i+t)$ ensures us that these case can appear only a finite number of times.
\end{enumerate}

\item Consider now that $buf_{r}(i+t)=(m',r',q',d',c')$.\\
Assume that $q'=p_{t}$ and $m'=m$, then  $buf_{r}(i+t)$ belongs to $C_{m}$ (the type of $C_{m}$ is then identical to the one of  $buf_{r}(i+t)$). Consequently, we have a contradiction with the definition of $C_{m}$. This implies that $q'\neq p_{t}$ or $m'\neq m$. Let $C_{m'}$ be the caterpillar whose $buf_{r}(i+t)$ belongs. Consider the three possible cases:

\begin{enumerate}
\item $C_{m'}$ is of type $S$: we can apply the induction hypothesis to $C_{m'}$ since its head stays in a buffer of rank greater or equals to $i+t$. Consequently, $C_{m'}$ becomes a caterpillar of type $F$ or $A$ in a finite time. That leads us to one of the following cases.
\item $C_{m'}$ is of type $A$: following Lemmas \ref{lem:prelem1} and \ref{lem:prelem2}, $C_{m'}$ disappears in a finite time. Then, $buf_{r}(i+t)$ becomes empty. That leads us to point 1.
\item $C_{m'}$ is of type $F$:  following Lemmas \ref{lem:prelem1} and \ref{lem:prelem3}, $C_{m'}$ disappears or becomes a caterpillar of type $S$ and length $1$ in a finite time. In all cases, $buf_{r}(i+t)$ becomes empty (since $i+t=D+1\geq2$). That leads us to point 1.
\end{enumerate}
\end{enumerate}

\item [Case 2.2:] $2\leq i+t\leq D$.\\
Consider the following cases:

\begin{enumerate}
\item $buf_{r}(i+t)=\varepsilon$.\\
Assume w.l.g. that $s=choice_{r}(i+t)$ and $buf_{s}(i+t-1)=(m',r,q',d',c')$.
By the construction of rule $\boldsymbol{(R_{8})}$ and the definition of a caterpillar, $r$ is enabled if and only if $buf_{nextHop_{r}(d')}(i+t+1)$ is not the tail of an abnormal caterpillar $C_{m'}$ associated to $m'$. Let us study the following cases:

\begin{enumerate}
\item $C_{m'}$ is of type $S$:  we can apply the induction hypothesis to $C_{m'}$ since its head stays in a buffer of rank greater or equals to $i+t+1$. Consequently, $C_{m'}$ becomes a caterpillar of type $F$ or $A$ in a finite time. That leads us to one of the following cases.
\item $C_{m'}$ is of type $A$: following Lemma \ref{lem:prelem1}, $C_{m'}$ disappears in a finite time. Then, $buf_{nextHop_{r}(d')}(i+t+1)$ becomes empty.
\item $C_{m'}$ is of type $F$:  following Lemma \ref{lem:prelem1}, $C_{m'}$ disappears in a finite time (it cannot become a caterpillar of type $S$ and length $1$ since $buf_{r}(i+t)=\varepsilon$). Consequently, $buf_{nextHop_{r}(d')}(i+t+1)$ becomes empty in a finite time.
\end{enumerate}

Then, Rule $\boldsymbol{(R_{8})}$ is enabled for $r$ in a finite time. This rule remains infinitely enabled since no message of type $(m'',r',r,d'',c'')$ can be copied in $buf_{nextHop_{r}(d')}(i+t+1)$ (indeed, the contrary implies that $nextHop_{r}(d')$ executes rule $\boldsymbol{(R_{8})}$ whereas $buf_{r}(i+t)=\varepsilon$). Since the daemon is weakly fair, $r$  executes rule $\boldsymbol{(R_{8})}$ in a finite time. The result of this execution is one of the following:

\begin{enumerate}
\item If $choice_{r}(i+t)=p_{t}$, then the head of $C_{m}$ goes up of one buffer when $r$ executes rule $\boldsymbol{(R_{8})}$.
\item If $choice_{r}(i+t)=s\neq p_{t}$, then $buf_{r}(i+t)$ takes the value $(m',r',s,d',c)$ when $r$ executes rule $\boldsymbol{(R_{8})}$. This situation is similar to the one of point 2 below. Note that the fairness of  $choice_{r}(i+t)$ ensures us that these case can appear only a finite number of times.
\end{enumerate}

\item If $buf_{r}(i+t)=(m',r',q',d',c')$, the reasoning is similar to the one of point 2 of case 2.1. Consequently, that leads us to point 1 in a finite time.
\end{enumerate}

In conclusion of case 2 ($p_{t}\neq d$), the head of $C_{m}$ goes up of one buffer in a finite time. Then, the induction hypothesis allows us to state that $C_{m}$ becomes a caterpillar of type $F$ or $A$ in a finite time. Consequently, $\boldsymbol{(P_{l})}$ is satisfied. 
\end{description}
\end{description}
\end{description}
\end{proof}

\paragraph{Snap-stabilization when routing tables are correct in the initial configuration.} Now, we assume that routing tables are correct in the initial configurations and we prove that \AD is a snap-stabilizing algorithm for specification $\mathcal{SP}'$.

\begin{lemma} \label{lem:chenilleSA}
Let $\gamma$ be a configuration in which routing tables are correct  and $m$ be a message of Destination $d$ existing in $\gamma$. Under a weakly fair daemon, every normal caterpillar of type $S$ associated to $m$ becomes a caterpillar of type $A$ in a finite time.
\end{lemma}

\begin{proof}
Let $\gamma$ be a configuration of the network in which routing tables are correct and $m$ be a message (of Destination $d$) existing in $\gamma$. Let $C_{m}=buf_{p_{1}}(1)...buf_{p_{t}}(t)$ ($t\geq1$) be a normal caterpillar of type $S$ associated to $m$.

By Lemma \ref{lem:prelem4}, $C_{m}$ becomes a caterpillar of type $A$ or $F$ in a finite time. In the first case, the proof ends here. In the second case (which is possible if $D+1-t\leq d(p_{t},d)$ in $\gamma$), it follows by Lemma \ref{lem:prelem3} that $C_{m}$ becomes a caterpillar of type $S$ of length $1$ in a finite time. Then, we have: $C_{m}=buf_{p_{1}}(1)$. 

Following Lemma \ref{lem:prelem4}, $C_{m}$ becomes a caterpillar of type $F$ or $A$ in a finite time. Assume that $C_{m}$ becomes a caterpillar of type $F$. This implies that $m$ have been forwarded $D$ times without reach its destination. This result is absurd since we have by definition that $dist(p_{1},d)\leq D$ and we assumed that routing tables are correct and constant. Consequently, $C_{m}$ becomes a caterpillar of type $A$ in a finite time.
\end{proof}

\begin{lemma} \label{lem:depotD}
If routing tables are correct, every processor can generate a first message (\emph{i.e.} it can execute $\boldsymbol{(R_{1})}$) in a finite time under a weakly fair daemon .
\end{lemma}

\begin{proof}
Let $p$ be a processor of the network which have a message $m$ (of Destination $d$) to forward. As $p$ have a waiting message, the higher layer put $request_{p}=true$ whatever its value in the initial configuration.

Assume that $buf_{p}(1)$ already contains a message. Let $C_{m}$ be the caterpillar which contains this buffer. We must distinguish the following cases:

\begin{description}
\item [Case 1:] $C_{m}$ is of type $F$. Following Lemma \ref{lem:prelem3}, $C_{m}$ becomes a caterpillar of type $S$ in a finite time. That leads us to case 2.
\item [Case 2:] $C_{m}$ is of type $S$. Following Lemma \ref{lem:chenilleSA}, $C_{m}$ becomes a caterpillar of type $A$ in a finite time. That leads us to case 3.
\item [Case 3:] $C_{m}$ is of type $A$. Following Lemma \ref{lem:prelem2}, $C_{m}$ disappears in a finite time.
\end{description}

In all cases, we obtain that $buf_{p}(1)$ becomes empty in a finite time. It remains empty while $p$ does not execute rule $\boldsymbol{(R_{1})}$ (since it is the only rule which can put a message in this buffer). In these case, $\boldsymbol{(R_{1})}$ is enabled for $p$ if and only if $buf_{nextHop_{p}(d)}(2)\neq(m,r',p,d,c)$. 

Assume that this condition is not satisfied. This implies (by definition of a caterpillar) that $buf_{nextHop_{p}(d)}(2)$ is the tail of an abnormal caterpillar $C'_{m}$. Following sequentially Lemmas \ref{lem:prelem4} and \ref{lem:prelem1}, $C'_{m}$ disappear in a finite time (note that the merge with $buf_{p}(1)$ is impossible since this buffer is empty). Moreover, $buf_{nextHop_{p}(d)}(2)$ can not be fill by a message of type $(m,r',p,d,c)$ (since $buf_{p}(1)$ is empty). Consequently, rule $\boldsymbol{(R_{1})}$ is infinitely enabled for Processor $p$. As the daemon is weakly fair, $p$ executes this rule in a finite time, that leads to the lemma.
\end{proof}

\begin{lemma} \label{lem:transportD}
If a message $m$ is generated by \AD in a configuration in which routing tables are correct, \AD delivers $m$ to its destination in a finite time under a weakly fair daemon.
\end{lemma}

\begin{proof}
The generation of a message $m$ (of Destination $d$) by \AD results from the execution of rule $\boldsymbol{(R_{1})}$ by the processor which sends $m$. This rule creates a normal caterpillar of type $S$ associated to $m$. Following Lemma \ref{lem:chenilleSA}, this caterpillar becomes a caterpillar of type $A$ in a finite time. It is due to the execution of rule $\boldsymbol{(R_{4})}$ or $\boldsymbol{(R_{10})}$ by $d$. These rules delivers the message to the higher layer of $d$, that ends the proof.
\end{proof}

\begin{proposition} \label{prop:snapTRD}
\AD is a snap-stabilizing message forwarding protocol for $\mathcal{SP}'$ if routing tables are correct in the initial configuration.
\end{proposition}

\begin{proof}
Assume that routing tables are correct in the initial configuration. To prove that \AD is a snap-stabilizing message forwarding protocol for specification $\mathcal{SP}'$, we must prove that :

\begin{enumerate}
\item If a processor $p$ requests to send a message, then the protocol is initiated by at least one starting action on $p$ in a finite time. In our case, the starting action is the execution of $\boldsymbol{(R_{1})}$. Lemma \ref{lem:depotD} proves this property. 
\item After a starting action, the protocol is executed according to $\mathcal{SP}'$. If we consider that $\boldsymbol{(R_{1})}$ have been executed at least one time, we can prove that:

\begin{itemize}
\item The first property of $\mathcal{SP}'$ is always satisfied (following Lemma \ref{lem:depotD} and the fact that the waiting for the sending of new messages is blocking). 
\item The second property of $\mathcal{SP}'$ is always satisfied (following Lemma \ref{lem:transportD}). 
\end{itemize}

\end{enumerate}

Consequently, we deduce the proposition.
\end{proof}

\paragraph{Self-stabilization.} Now, we assume that routing tables are corrupted in the initial configurations and we prove that \AD is a self-stabilizing algorithm for specification $\mathcal{SP}'$.

\begin{proposition} \label{prop:selfD}
\AD is a self-stabilizing message forwarding protocol for $\mathcal{SP}'$ even if routing tables are corrupted in the initial configuration when $\mathcal{A}$ runs simultaneously.
\end{proposition}

\begin{proof}
Remind that $\mathcal{A}$ is a self-stabilizing silent algorithm for computing routing tables running simultaneously to \AD. Moreover, we assumed that $\mathcal{A}$ has priority over \AD (\emph{i.e.} a processor which have enabled actions for both algorithms always chooses the action of $\mathcal{A}$). This guarantees us that routing tables are correct and constant in a finite time regardless of their initial states. 

By Proposition \ref{prop:snapTRD}, \AD is a snap-stabilizing message forwarding protocol for specification $\mathcal{SP}'$ when it starts from a such configuration. Consequently, we can conclude on the proposition.
\end{proof}

\paragraph{Snap-stabilization.} We still assume that routing tables are corrupted in the initial configuration and we prove that \AD is a snap-stabilizing algorithm for specification $\mathcal{SP}$.

\begin{lemma} \label{lem:perteD}
Under a weakly fair daemon, \AD does not delete a valid message without delivering it to its destination even if $\mathcal{A}$ runs simultaneously.
\end{lemma}

\begin{proof}
When \AD accepts a new valid message $m$, the processor which sends $m$ executes rule $\boldsymbol{(R_{1})}$. By construction of the rule, this execution creates a normal caterpillar $C_{m}$ of type $S$ associated to $m$.

While $m$ is not delivered to its destination, we know, by Lemmas \ref{lem:prelem4} and \ref{lem:prelem3}, that $C_{m}$ follows infinitely often the above cycle:

\begin{itemize}
\item $C_{m}$ is of type $S$ and becomes of type $F$ (type $A$ is impossible since $m$ is not delivered).
\item $C_{m}$ is of type $F$ and becomes of type $S$.
\end{itemize}

This implies that there always exists at least one copy of $m$ in $buf_{p}(1)$ (if $p$ is the sending processor of $m$). Then, this message is not deleted without being delivered to its destination.
\end{proof}

\begin{lemma} \label{lem:duplicationD}
Under a weakly fair daemon, \AD never duplicates a valid message even if $\mathcal{A}$ works simultaneously.
\end{lemma}

\begin{proof}
It is obvious that the emission of a message $m$ by rule $\boldsymbol{(R_{1})}$ only creates one caterpillar of type $S$ associated to $m$. 

Then, observe that all rules are designed to obtain the following property: if a caterpillar has one head in a configuration, it also has one head in the following configuration whatever rules have been applied. Indeed, this property is ensured by the fact that the next processor on the path of a message $m$ is computed (and put in the second field on the message) when $m$ is copied into a buffer $buf_{p}(i)$ (not when it is forwarded to a neighbor). Consequently, if there is a routing table move after the copy of $m$ in $buf_{p}(i)$, the caterpillar does not fork. The head of the caterpillar remains unique.

We can conclude that, for any valid message $m$, there always exists a unique caterpillar $C_{m}$ associated to $m$. Assume that $m$ is delivered. By construction of rules $\boldsymbol{(R_{4})}$ and $\boldsymbol{(R_{10})}$, $C_{m}$ becomes of type $A$. Following Lemma \ref{lem:prelem2}, $C_{m}$ disappears in a finite time. Consequently, $m$ cannot be delivered several times.
\end{proof}

\begin{theorem} \label{th:snapD}
\AD is a snap-stabilizing message forwarding protocol for $\mathcal{SP}$ even if routing tables are corrupted in the initial configuration when $\mathcal{A}$ runs simultaneously.
\end{theorem}

\begin{proof}
Proposition \ref{prop:selfD} and Lemma \ref{lem:perteD} allows us to conclude that \AD is a snap-stabilizing message forwarding protocol for specification $\mathcal{SP}'$ even if routing tables are corrupted in the initial configuration on condition that $\mathcal{A}$ runs simultaneously.

Then, using this remark and Lemma \ref{lem:duplicationD}, we obtain the result.
\end{proof}

\subsection{Time complexities} \label{sub:analyseD}

Since our algorithm needs a weakly fair daemon, there is no points to do an analysis in terms of steps. It is why all the following complexities analysis are given in rounds. Let $R_{\mathcal{A}}$ be the stabilization time of $\mathcal{A}$ in terms of rounds.

In order to lighten this paper, we present only key ideas of this section proofs.

\begin{proposition} \label{prop:analysemesD}
In the worst case, $\Theta(nD)$ invalid messages are delivered to Processor $d$.
\end{proposition}

\begin{sketchproof}
In the initial configuration, the system has at most $n(D+1)$ distinct invalid messages of Destination $d$. Then, the number of invalid messages deliver to $d$ is in $O(nD)$. 

We can obtain the lower boundwith a chain of $n=2q+1$ processors labeled $p_{1},p_{2},...,p_{n}$. Assume that all buffers of rank least or equals to $q+1$ initially contain a message of destination $p_{q+1}$ and other buffers are empty. Moreover, assume that routing tables are initially correct. Then, \AD delivers all invalid messages of this initial configuration to $p_{q+1}$. This initial configuration contains $n(q+1)=n(\frac{D}{2}+1)\in\Theta(nD)$ invalid messages. The result follows.
\end{sketchproof}

\begin{proposition} \label{prop:complexiteD}
In the worst case, a message $m$ (of Destination $d$) needs $O(max(R_{\mathcal{A}},nD\Delta^{D}))$ rounds to be delivered to $d$ once it has been sent out by its source.
\end{proposition}

\begin{sketchproof}
In a first time, one must prove by induction the following fact: if $\gamma$ is a configuration in which routing tables are correct and in which a message of Destination $d$ exists and $C_{m}$ is a caterpillar of type $S$ associated to $m$ which head is a buffer of rank $1\leq i+t-1<D+1$ on $p\neq d$, then the head of $C_{m}$ goes up of one buffer in at most $O(\Delta^{D+1-(i+t-1)})$ round if there exists no abnormal caterpillar whose tail is a buffer of rank greater than $i+t$.

In a second time, it is possible to show that $\mathcal{C}$, the set of abnormal caterpillars in $\gamma$ looses at least one element during the $O(\Delta^{D})$ rounds which follow $\gamma$. Then, we can say that, when routing tables are correct, an accepted message is forwarded in at most $O(nD\Delta^{D})$ rounds.

Finally, we can deduce the result when $m$ is emitted in a configuration in which routing tables are not correct since the message is delivered in at most $O(nD\Delta^{D})$ rounds after routing tables computation (which takes at most $O(R_{\mathcal{A}})$ rounds if $m$ is not delivered during the routing tables computation since we have assumed the priority of $\mathcal{A}$).
\end{sketchproof}

\begin{proposition} \label{prop:delaiD}
The delay (waiting time before the first emission) and the waiting time (between two consecutive emissions) of \AD is $O(max(R_{\mathcal{A}},nD\Delta^{D}))$ rounds in the worst case.
\end{proposition}

\begin{sketchproof}
Let $p$ be a processor which has a message of Destination $d$ to emit. By the fairness of $choice_{p}(d)$, we can say that $m$ is sent after at most $(\Delta-1)$ releases of $buf_{p}(1)$. The result of Proposition \ref{prop:complexiteD} allows us to say that $buf_{p}(1)$ is released in $O(max(R_{\mathcal{A}},nD\Delta^{D}))$ rounds at worst. Indeed, we can deduce the result.
\end{sketchproof}

The complexity obtained in Proposition \ref{prop:complexiteD} is due to the fact that the system delivers a huge quantity of messages during the forwarding of the considered message. It's why we interest now in the amortized complexity (in rounds) of our algorithm. For an execution $\Gamma$, this measure is equal to the number of rounds of $\Gamma$ divided by the number of delivered messages during $\Gamma$ (see \cite{CLRS02} for a formal definition).

\begin{proposition} \label{prop:amortieD}
The amortized complexity (to forward a message) of \AD is in $O(max$ $(R_{\mathcal{A}},D))$ rounds when there exists no invalid messages.
\end{proposition}

\begin{sketchproof}
In a first time, we must prove the following property: if $\gamma$ is a configuration in which at least one caterpillar of type $S$  is present, routing tables are correct, and there exists no invalid messages, then \AD delivers at least one message to a processor in the $3D+1$ rounds following $\gamma$.

Assume now an initial configuration in which routing tables are correct and in which there exists no invalid messages. Let $\Gamma$ be one execution which leads to the worst amortized complexity. Let $R_{\Gamma}$ be the number of rounds of $\Gamma$. By the last remark, we can say that \AD delivers at
least $\frac{R_{\Gamma}}{3D+1}$ messages during $\Gamma$. So, we have an amortized complexity of $\frac{R_{\Gamma}}{\frac{R_{\Gamma}}{3D+1}}\in\Theta(D)$. Then, the announced result is obvious.
\end{sketchproof}

\subsection{Conclusion}

In this section, we prove that we can adapt the ``distance-based'' deadlock-free controller defined in \cite{MS78} to obtain a snap-stabilizing message forwarding algorithm. Our algorithm is mainly based on an acknowledgement scheme. Each message is re-emitted until it reaches its destination. As routing tables stabilize in a finite time, we are ensured that, in the worst case, the message is re-emitted after the end of computation of routing tables. Hence, it can reach its destination normally.

The initial fault-free protocol uses $n(D+1)$ buffers for the whole network and our protocol uses exactly the same number of buffers. Consequently, our protocol ensures a stronger safety and fault-tolerance with respect the initial one without overcost in space. Our time analysis (see Section \ref{sub:analyseD}) shows that this stronger safety does not leads to an overcost in time.

\section{Conclusion}\label{sec:Conclusion}

In this paper, we provide the first algorithms (at our knowledge) to solve the message forwarding problem in a snap-stabilizing way (when a self-stabilizing algorithm for computing routing tables runs simultaneously) for a specification which forbids message losses and duplication. This property implies the following fact: our protocol can forward any emitted message to its destination regardless of the state of routing tables in the initial configuration. Such an algorithm allows the processors of the network to send messages to other without waiting for the routing table computation. We use a tool called ``buffer graph'' which has been introduced in \cite{MS78}. This paper proposed an adaptation of two "buffer graphs" in order to control the effect of routing table moves on messages. Our analysis shows that we ensure snap-stabilization without significant overcost in space or in time with respect to the fault-free algorithm.

\cite{MS78} also proposed other buffer graphs. So, it is natural to wonder if they could be adapted to tolerate transient faults. In particular, one of them (based on the acyclic covering of the network, see also \cite{T01}) is very interesting since it needs less buffers per processor in general (3 for a ring, 2 for a tree...). But, authors of \cite{KR07} show that it is NP-hard to compute the size of the acyclic covering of any graph. So, this buffer graph cannot be easily applied to any network. A very important open problem is the following: what is the minimal number of buffers per processor to allow snap-stabilization on the message forwarding problem ?

Another way to improve our protocol is to speed up the message forwarding in the worst case (without increasing amortized complexity). In this goal, we believe that we can keep our protocol and modify the fair scheme of selection of messages $choice_{p}(d)$. In fact, the complexity of our algorithm depends on the number of messages which can ``pass'' a specific message at each hop.

Our protocol has the following drawback: when a message $m$ is delivered to a processor $p$, $p$ cannot determine if $m$ is valid or not. This can bring some problems for applications which use these messages. So, an interesting way of future researches could be to design a protocol which solves this problem. In \cite{CDV06} the authors propose an efficient solution for the PIF problem that deals with a similar problem, unfortunately their approach does not seem suitable for our problem.

Finally, it would be interesting to carry our protocol in the message passing model (a more realistic model of distributed system) in order to enable snap-stabilizing message forwarding in a real network. To our knowledge, in this model, only two snap-stabilizing protocols exist in the literature (\cite{DDNT08,DT06}). The problem to carry automatically a protocol from the state model to the message passing model is still open. 


\newpage

\bibliographystyle{plain}
\bibliography{Biblio}

\end{document}